\documentclass[12pt]{article}
 \pdfoutput=1
\usepackage{amssymb}
\usepackage{amsmath}
\usepackage{amstext}
\usepackage{graphicx,epsfig}
\usepackage{epsfig}
\usepackage{verbatim} 
\usepackage{fancybox}
\usepackage{color}
\usepackage{ulem}
\usepackage{enumitem}
\usepackage{subfigure}
\usepackage{bbm}
\usepackage{parskip}
\usepackage{cite}

\newcommand{\Comment}[1]{{}}
\definecolor{MyDarkBlue}{rgb}{0.15,0.15,0.45}
\usepackage[linktocpage=true]{hyperref}
\hypersetup{
colorlinks=true,
citecolor=MyDarkBlue,
linkcolor=MyDarkBlue,
urlcolor=MyDarkBlue,
pdfauthor={Kurt Hinterbichler},
pdftitle={Ghost-Free Derivative Interactions for a Massive Graviton},
pdfsubject={hep-th}
}

\newcommand{\be}{\begin{equation}}
\newcommand{\ee}{\end{equation}}
\newcommand{\bea}{\begin{eqnarray}}
\newcommand{\eea}{\end{eqnarray}}
\newcommand{\beas}{\begin{eqnarray*}}
\newcommand{\eeas}{\end{eqnarray*}}
\newcommand{\nn}{\nonumber}
\newcommand{\stu}{St\"uckelberg\ }

\def\({\left(}
\def\){\right)}

\newcommand{\Tr}{\text{Tr}}

\newcommand{\la}{\langle}
\newcommand{\ra}{\rangle}

\newcommand{\half}{\frac{1}{2}}

\newcommand{\beq}{\begin{equation}}
\newcommand{\eeq}{\end{equation}}
\newcommand{\bal}{\begin{aligned}}
\newcommand{\eal}{\end{aligned}}

\newlength{\fdagwidth}
\newlength{\diagupwidth}
\newlength{\stepback}
\newcommand{\fdag}[2][\diagup]{\text{$#2$\settowidth{\fdagwidth}{$#2$}\settowidth{\diagupwidth}{$#1$}\setlength{\stepback}{0.5\fdagwidth}\hspace{-\stepback}\hspace{-0.5\diagupwidth}$#1$\hspace{\stepback}\hspace{-0.5\diagupwidth}}}

\numberwithin{equation}{section}

\begin{document}


\begin{center}
{\LARGE \bf{ Supersymmetric Partially Massless Fields and }}\\ \vspace{.2cm}
{\LARGE \bf{Non-Unitary Superconformal Representations}}
\end{center} 

\vspace{2truecm}

\thispagestyle{empty}
\centerline{{\Large Sebastian Garcia-Saenz,${}^{\rm a,}$\footnote{\href{mailto:sebastian.garcia-saenz@iap.fr}{\texttt{sebastian.garcia-saenz@iap.fr}}} Kurt Hinterbichler,${}^{\rm b,}$\footnote{\href{mailto:kurt.hinterbichler@case.edu}{\texttt{kurt.hinterbichler@case.edu}}} Rachel A. Rosen${}^{\rm c,}$\footnote{\href{mailto:rar2172@columbia.edu}{\texttt{rar2172@columbia.edu}}}}}
\vspace{.7cm}

 \centerline{{\it ${}^{\rm a}$Sorbonne Universit\'e, UPMC Univ.\ Paris 6 and CNRS, UMR 7095,}}
 \centerline{{\it Institut d'Astrophysique de Paris, GReCO, 98bis boulevard Arago, 75014 Paris, France}} 
 \vspace{.35cm}

 \centerline{{\it ${}^{\rm b}$CERCA, Department of Physics, Case Western Reserve University, }}
 \centerline{{\it 10900 Euclid Ave, Cleveland, OH 44106}} 
 \vspace{.35cm}

 \centerline{{\it ${}^{\rm c}$Center for Theoretical Physics, Department of Physics, }}
 \centerline{{\it Columbia University, New York, NY 10027}} 
 \vspace{.35cm}

\begin{abstract}
\vspace{-0.5truecm}
We find and classify the ${\cal N}=1$ SUSY multiplets on AdS$_4$ which contain partially massless fields.  We do this by studying the non-unitary representations of the $d=3$ superconformal algebra of the boundary.  The simplest super-multiplet which contains a partially massless spin-2 particle also contains a massless photon, a massless spin-$3/2$ particle and a massive spin-$3/2$ particle.  The gauge parameters form a Wess-Zumino super-multiplet which contains the gauge parameters of the photon, the partially massless graviton, and the massless spin-$3/2$ particle.  We find the AdS$_4$ action and SUSY transformations for this multiplet.   More generally, we classify new types of shortening conditions that can arise for non-unitary representations of the $d=3$ superconformal algebra.

\end{abstract}

\newpage

\setcounter{tocdepth}{2}
\tableofcontents
\newpage

\section{Introduction}
\parskip=5pt
\normalsize

De Sitter (dS) and anti de Sitter (AdS) spacetimes allow for exotic irreducible representations which go by the name of partially massless (PM) particles \cite{Deser:1983tm,Deser:1983mm,Higuchi:1986py,Brink:2000ag,Deser:2001pe,Deser:2001us,Deser:2001wx,Deser:2001xr,Zinoviev:2001dt,Skvortsov:2006at,Skvortsov:2009zu}.   A partially massless particle of spin $s$ possesses a mass and also gauge symmetry, labelled by a depth $t \in \{0, 1, . . . , s - 1\} $.  The gauge symmetry eliminates the helicity components $0, 1, \ldots , t$ from the particle, leaving a number of degrees of freedom intermediate between that of a massless and a massive field.  These gauge invariances emerge at fixed values of the particle mass relative to the (A)dS curvature.  

In this work, we consider the supersymmetric (SUSY) extension of partially massless representations on AdS$_4$.  We do this through the AdS/CFT correspondence by studying the representations of the $d=3$ superconformal algebra of the boundary.   In AdS, the partially massless representations are non-unitary, and so we must study the non-unitary representations of the superconformal algebra.  Unitary superconformal representations in three dimensions have been extensively studied \cite{Dolan:2008vc,Bhattacharya:2008zy,Cordova:2016emh}, but the non-unitary case has remained relatively unexplored (see however \cite{Oshima:2016gqy,Sen:2018del}).  

In the boundary CFT, partially massless particles correspond to short multiplets with conformal dimension $\Delta=t+2$, which have a null descendent at level $s-t$.  We will find that partially massless particles appear in {superconformal} multiplets which contain the standard four conformal primaries for generic massive particles.  For instance, the simplest super-multiplet which contains a partially massless spin-2 particle also contains a massless photon, a massless spin-$3/2$ particle and a massive spin-$3/2$ particle.  The scalar partially massless gauge parameter is a Wess-Zumino super-multiplet which contains the scalar gauge parameters of the photon and the partially massless graviton, as well as the spin-$1/2$ gauge parameter of the massless spin-$3/2$ particle.  We explicitly find the AdS action and SUSY transformations for this multiplet.   

Away from the partially massless values, we classify new types of BPS-like shortening conditions that can arise for non-unitary representations of the $d=3$ superconformal algebra.  For fields that obey the usual ``standard quantization" we recover the expected short multiplets whose duals consist of massless fields in AdS$_4$.  We also find a shortening condition for ``alternatively quantized" fields which themselves have finite dimensional Verma modules, including an additional shortening condition for certain alternatively quantized partially massless fields.  Finally, for $\Delta = 1$, we find a supersymmetric example of the exotic ``extended modules'' of \cite{Brust:2016gjy}.

Studies of massive supergravity can be found in \cite{Malaeb:2013lia,Malaeb:2013nra,DelMonte:2016czb,Ondo:2016cdv,Zinoviev:2018juc}.  Partially massless supergravity is potentially interesting for a number of reasons.  While theories of free partially massless particles are straightforward to construct, various studies and no-go theorems seem to forbid consistent theories of a single {interacting} partially massless spin-2 particle~\cite{Zinoviev:2006im,Hassan:2012gz,Hassan:2012rq,deRham:2012kf,Hassan:2013pca,Deser:2013uy,deRham:2013wv,Zinoviev:2014zka,Garcia-Saenz:2014cwa,Hinterbichler:2014xga,Joung:2014aba,Alexandrov:2014oda,Hassan:2015tba,Hinterbichler:2015nua,Cherney:2015jxp,Gwak:2015vfb,Gwak:2015jdo,Garcia-Saenz:2015mqi,Hinterbichler:2016fgl,Bonifacio:2016blz,Apolo:2016ort,Apolo:2016vkn,Bernard:2017tcg,Boulanger:2018dau}.  However, it may still be possible to construct theories of partially massless particles interacting with other fields, and there are examples of Vasiliev-like theories with infinite towers of fields \cite{Bekaert:2013zya,Basile:2014wua,Alkalaev:2014nsa,Joung:2015jza,Brust:2016zns}.  A theory of an interacting supersymmetric partially massless multiplet would not contradict any known no-go theorems.  Finally, non-unitary conformal theories and representations have found applications in a range of condensed matter systems \cite{Maassarani:1996jn}, and in understanding the analytic structure of the conformal blocks in unitary theories \cite{Penedones:2015aga}.  It is thus useful to explore and categorize the non-unitary irreducible representations of the superconformal group.

\bigskip

{\bf Conventions}:
We work in $D=4$ spacetime dimensions for AdS, and $d=3$ Euclidean spatial dimensions for CFT.  
We use the mostly plus metric signature convention.  Tensors are symmetrized and anti-symmetrized with unit weight, i.e $T_{(\mu\nu)}=\half \left(T_{\mu\nu}+T_{\nu\mu}\right)$,   $T_{[\mu\nu]}=\half \left(T_{\mu\nu}-T_{\nu\mu}\right)$.  We use $\mu,\nu,\cdots$ for AdS spacetime indices, $i,j,\cdots$ for flat CFT 3d tensor indices, $a,b,\cdots$ for 3d CFT spinor indices.   Conventions for 3d spinor quantities are detailed in Appendix \ref{spinorappendix}, and conventions for 4D spinor quantities are detailed in Appendix \ref{app:ads conventions}.  Conventions for the curvature tensors and covariant derivatives are those of Carroll \cite{Carroll:2004st}.  $L$ the AdS radius.  The scalar curvature $R$ and cosmological constant $\Lambda$ are related as $R=-{12\over {L}^2},\ \Lambda=-{3\over {L}^2}.$ 

\section{Partially Massless Fields and Their Dual Operators}

Through the AdS/CFT correspondence, a spin $s$ field in ${\rm AdS}_4$ corresponds to a spin $s$ primary operator in ${\rm CFT}_3$.  The mass $m$ of the field and the dimension $\Delta$ of the primary are related by
\be m^2L^2=\begin{cases} \Delta(\Delta-3), & s=0\, ,\\ \left(\Delta+s-2\right)\left(\Delta-s-1\right), &  s\geq 1/2\, .  \end{cases}   \label{AdSCFTformpe} \ee
For a given mass, there are two different ways to quantize the field in AdS$_4$.   These correspond to the greater and lesser roots of \eqref{AdSCFTformpe}, $\Delta_{\pm}$.  $\Delta_+$ corresponds to the so-called ``standard quantization'' which covers the operators with $\Delta>3/2$, and $\Delta_-$ to the ``alternate quantization" \cite{Klebanov:1999tb} which covers the operators with $\Delta<3/2$.  

The unitarity bound for primary operators is \cite{Mack:1975je,Jantzen1977,Minwalla:1997ka}
\be \Delta\geq  \begin{cases}  1/2, & s=0\,,\\ 1, & s=1/2\, , \\ s+1, & s\geq 1\, .\end{cases}\label{unitboundd}\ee
Theories containing primary operators violating these bounds are necessarily non-unitary.

\subsection{Bosons}

Bosonic fields of spin $s$ on AdS$_4$ are carried by fully symmetric tensors $\phi_{\mu_1\ldots \mu_s}$ whose equations of motion can be brought to the form:
\bea && \left[\square -{1\over L^2}\left(s(s-2)-2\right)-m^2\right]\phi_{\mu_1\ldots \mu_s}=0, \label{eomexp} \nn\\
&&  \nabla^{\nu}\phi_{\nu\mu_2\ldots \mu_s}=0, \label{transverseexp} \nn\\
&& \phi^\nu_{\phantom{\nu} \nu\mu_3\ldots\mu_s}=0,\label{massivesys}\eea
i.e.  they are transverse and traceless and obey the wave equation.  Here $\square\equiv\nabla^\mu \nabla_\mu$ is the bare curved space Laplacian.

For bosons of spin $s\geq 1$, partially massless points occur at the following mass values labelled by $t$,
\be m_{s,t}^2=-{1\over L^2}\left(s-t-1\right)\left(s+t\right),\ \ \ t=0,1,2,\cdots,s-1\, .\label{pmvalues}\ee
Here, $t$ is known as the {\it depth} of partial masslessness of the field.   The massless field is the one at depth $t=s-1$.  At these mass values, the equations \eqref{massivesys} becomes invariant under a partially massless gauge symmetry,
\be \delta \phi_{\mu_1\ldots\mu_s}=\nabla_{(\mu_{t+1}}\cdots \nabla_{\mu_{s}}\xi_{\mu_1\ldots \mu_t)}+\ldots \label{PMgaugesym}
\ee
with a symmetric tensor gauge parameter $\xi_{\mu_1\ldots \mu_t}$, which on shell is subject to its own conditions analogous to \eqref{massivesys}.   The $\ldots$ in \eqref{PMgaugesym} stand for lower-derivative terms proportional to factors of the curvature $1/L^2$.  Note that the depth $t$ corresponds to the number of indices on the gauge parameter in \eqref{PMgaugesym}.

The partially massless values \eqref{pmvalues} correspond to the conformal dimensions
\be \Delta_{s,t}=t+2\, . \label{confdimbse} \ee
(This is the normal quantization root $\Delta_+$, the alternate root is $\Delta_-=1-t$.)
They violate the unitarity bound \eqref{unitboundd} except for the massless case at $t=s-1$, which saturates it.

A primary state $|\Delta\ra^{i_1\cdots i_s}$ with the conformal dimension \eqref{confdimbse} gives rise to a short multiplet \cite{Dolan:2001ih}.  There is a null descendent at level $s-t$, 
\be P_{i_1}\ldots P_{i_{s-t}}  |\Delta\ra^{i_1\cdots i_s}=0\, ,\label{bosonmultiplyconsce}\ee
(this is the type II shortening in the language of \cite{Penedones:2015aga}).  This primary can be called a multiply-conserved current.

\subsection{Fermions}

A spin $s$ fermion on ${\rm AdS}_4$ is a totally symmetric tensor spinor $\Psi_{\mu_1\cdots \mu_{s-1/2}}$, whose equations of motion can be brought to the form: 
\bea && \left( i\fdag{{\nabla}}-\tilde m \right)\Psi_{\mu_1\cdots \mu_{s-1/2}}=0\, , \nn\\
&& {\nabla}^\nu \Psi_{\nu\mu_2\cdots \mu_{s-1/2}}=0 \, , \nn\\
&&  \gamma^\nu \Psi_{\nu\mu_2\cdots \mu_{s-1/2}}=0  \, , \label{fermioneqsyste}
\eea
i.e.  they are transverse, gamma traceless (hence traceless) and obey the Dirac equation.  Here ${\nabla}$ is the full AdS covariant derivative, which includes both Christoffel symbols acting on the tensor indices and a spin connection acting on the spinor index.  The parameter $\tilde m$ in the Dirac equation is related to the mass $m^2$ appearing in the AdS/CFT formula \eqref{AdSCFTformpe} by
\be \tilde m^2=\frac{(2 s-1)^2}{4 L^2}+m^2.\ee

For $s\geq3/2$, partially massless points occur at the following mass values labelled by the depth $t$,
\be \tilde m_{s,t}^2=\frac{(2 t+1)^2}{4 L^2}\,, \ \ \ m^2_{s,t}=\frac{(t-s+1) (s+t)}{L^2}\, ,\ \ \ t=1/2,3/2,\cdots,s-1\, .\label{PMvaluesfermionse}\ee
The massless field is the one at depth $t=s-1$.  At these mass values, the equations \eqref{fermioneqsyste} becomes invariant under a partially massless gauge symmetry,
\be \delta \Psi_{\mu_1\cdots \mu_{s-1/2}}={\nabla}_{(\mu_{t+1/2}}\cdots {\nabla}_{ \mu_{s-1/2}}\Phi_{\mu_1\ldots \mu_{t-1/2})}+\ldots \label{PMgaugesymf}
\ee
with a symmetric tensor spinor gauge parameter $\Phi_{\mu_1\ldots \mu_t}$. which on shell is subject to its own conditions analogous to \eqref{massivesys}.   The $\ldots$ in \eqref{PMgaugesymf} stand for lower-derivative terms proportional to factors of the curvature $1/L^2$.  Note that the depth $t$ corresponds to the spin of the gauge parameter in \eqref{PMgaugesymf}.

 The partially massless values \eqref{PMvaluesfermionse} have $m^2<0$, increasingly negative as $t$ decreases, with $m^2=0$ the massless value $t=s-1$.  However they have $\tilde m^2>0$.
They correspond to the conformal dimensions
\be \Delta_{s,t}=t+2\label{PMvaluesfermionscde}\, .\ee
(This is the normal quantization root $\Delta_+$, the alternate root is $\Delta_-=1-t$.)

The partially massless fields are dual to short multiplets with tensor spinor conformal primaries $|\Delta\ra^{i_1\cdots i_{s-{1/ 2}} \ a}$ ($a$ is the spinor index) that have a null descendent at level $s-t$, 
\be P_{i_1}\ldots P_{i_{s-t}} |\Delta\ra^{i_1\cdots i_{s-{1/ 2}} \ a}=0\, .\label{fermionmultiplyconsce}\ee
These are fermionic multiply-conserved current.

\section{$d=3$ Superconformal Representations}

We are interested in seeing how the partially massless bosons and fermions are joined together into supermultiplets.  For this we must study the SUSY representations of AdS$_4$, which are the same as the superconformal representations of the flat three dimensional boundary.
Unitary superconformal representations in three dimensions have been studied in \cite{Dolan:2008vc,Bhattacharya:2008zy}, and in other dimensions in \cite{Minwalla:1997ka,Cordova:2016emh}, and the multiplets on AdS$_4$ have also been studied \cite{deWit:1999ui}.  The partially massless values \eqref{confdimbse}, \eqref{PMvaluesfermionscde} fall below the unitarity bound \eqref{unitboundd}, so we will be interested in non-unitary superconformal representations, which have not been studied as far as we know.  Along the way, we will find new kinds of BPS-like shortening conditions for the non-unitary superconformal multiplets which have no analogue among the unitary representations.

\subsection{$d=3$ Superconformal Algebra}

The generators of the euclidean ${\cal N}=1$ superconformal algebra are 
\bea &&  P^i,\  J_{ij},\   D,\ \ K^i,  \    Q^a,\  S_a\,.
\eea 
Here $P^i$ are the translations and $J_{ij}$ the rotations, which together generate the Poincare transformations of 3d euclidean space.  $D$ is the dilation and $K^i$ the special conformal generators, which together with the Poincare generators generate the conformal symmetries of 3d euclidean space.  $Q^a$ is the supersymmetry, which together with the Poincare generators generate ${\cal N}=1$ SUSY.  $S_a$ is the special superconformal generator, which completes the SUSY generators and conformal generators into to the superconformal algebra.

We now list all the non-vanishing (anti)commutators of the 3d euclidean superconformal algebra \cite{Minwalla:1997ka} (see Appendix \ref{spinorappendix} for our conventions on 3d spinors and representations).
First there are the standard commutation relation of the Poincare algebra,
\bea \left[J^{ij},P^k\right] &=& i\left(-\delta^{ki} P^j+\delta^{kj} P^i\right)\,  ,  \nn \\
\left[J^{ij},J^{kl}\right]&=&i\left(-\delta^{ik}J^{jl}+\delta^{jk}J^{il}-\delta^{jl}J^{ik}+\delta^{il}J^{jk} \right)\,. \label{pvectorc}
\eea 
Next are the commutators which when taken together with \eqref{pvectorc} form the conformal algebra,
\bea \left[D,P^i\right]&=&  P^i \, ,  \label{weightp1} \nn\\ 
 \left[D,K^i\right]&=& - K^i \, ,\label{weightkm1}\nn\\
  \left[K^i,P^j\right]&=&2(\delta^{ij} D+iJ^{ij})\, ,\nn\\
   \left[J^{ij},K^k\right]&=& i\left(-\delta^{ki}K^j+\delta^{kj}K^i\right)\, .\label{kvectorc}
   \eea 
Then we have the commutators which when taken together with \eqref{pvectorc} form the ${\cal N}=1$ SUSY algebra,
\bea && \left\{Q^{a},Q^{b}\right\}=2\sigma_i^{ab}P^i\, , \nn \\ 
&& \left[ J_{ij},Q^{a}\right]=-{i\over 2}\left(\sigma_{ij}\right)^a_{\ b}Q^{b} \,. \label{mainsssyq}
\eea
The first line of \eqref{mainsssyq} is the fundamental anti-commutator of SUSY, the second indicates that $Q^a$ transforms as a spinor.

Finally we have the commutators which when taken with all the above fill out the superconformal algebra,
\bea && \left[ J_{ij},S^{a}\right]=-{i\over 2}\left(\sigma_{ij}\right)^a_{\ b}S^{b}\, ,\nn\\
&& \left[ D, Q^{a}\right]={1\over 2}Q^{a}\, ,\nn\\
&& \left[ D, S^{a}\right]=-{1\over 2}S^{a}\, ,\nn\\
&& \left\{S^{a},S^{b}\right\}=-2\sigma_i^{ab}K^i\,, \label{mainsssys}\nn\\
&&   \left\{Q^{a},S^{b}\right\}=2\epsilon^{ab}D-i\sigma_{ij}^{ab}J^{ij}\, ,\label{mainsccomae} \nn\\
&& \left[K_i,Q^{a}\right]=-\left(\sigma_i\right)^a_{\ b}S^{b}\, ,\nn\\
&&  \left[P_i,S^{a}\right]=\left(\sigma_i\right)^a_{\ b}Q^{b}\, .\label{remainingsuscyecae}
\eea
The first line of \eqref{remainingsuscyecae} indicates that $S^a$ transforms as a spinor, the second and third lines indicate that $Q^a$ carries dimension $1/2$ and $S^a$ carries dimension $-1/2$.

In radial quantization, the generators satisfy the following reality conditions\footnote{Note that with our conventions, outlined in Appendix \ref{spinorappendix}, there is a subtlety with the spinor indices; the condition ${Q^{a}}^\dag=S_{a}$ implies ${S_{a}}^\dag=Q^{a}$, but when raising and lowering indices we get a minus sign \eqref{raiseloweape}, ${Q_{a}}^\dag=-S^{a},\   {S^{a}}^\dag=-Q_{a}\,.$}
\be P_i^\dag=K_i,\ \ D^\dag=D,\ \ \ \ J_{ij}^\dag=J_{ij},\ \ \ {Q^{a}}^\dag=S_{a}\,. \label{PDJcong}\ee  It is straightforward to check that all the (anti)commutation relations above are consistent with the reality conditions \eqref{PDJcong}.

\subsection{Conformal representations}

Conformal representations are built by starting with a spin $s$ conformal primary operator of dimension $\Delta$.  We represent this as a state $|\Delta\ra^{a_1\cdots a_{2s}}$ carrying $2s$ completely symmetric spinor indices.  It transforms as an irreducible spin $s$ representation under rotations,
\bea && J_{ij}|\Delta\ra^{a_1\cdots a_{2s}}= -{i\over 2}\left(\sigma_{ij}\right)^{a_1}_{\ \ b} |\Delta\ra^{ba_2\cdots a_{2s}}-\cdots-{i\over 2}\left(\sigma_{ij}\right)^{a_{2s}}_{\ \ b} |\Delta\ra^{a_1\cdots a_{2s-1}b},\ \ \\
&& J_{ij}J^{ij} |\Delta\ra^{a_1\cdots a_{2s}}=2s(s+1)|\Delta\ra^{a_1\cdots a_{2s}}, \label{J2evprimconde}
\eea
and carries eigenvalue $\Delta$ under the dilation generator $D$,
\be D|\Delta\ra^{a_1\cdots a_{2s}}=\Delta |\Delta\ra^{a_1\cdots a_{2s}}.\label{Devprimconde}\ee
The primary has the property that it is annihilated by $K^i$,
\be K^i |\Delta\ra^{a_1\cdots a_{2s}}=0.\ee
The rest of the conformal representation is built by acting repeatedly with $P^i$ on the primary.  The representation is graded by the level, which is the number of $P$'s acting on the primary, so that at level $l$ the states are
\be P^{i_1}\cdots P^{i_l}|\Delta\ra^{a_1\cdots a_{2s}},\ee
and they have dimension $\Delta+l$.  Each level is then further decomposed into irreducible representations of $J_{ij}$.   This forms a Verma module where we go up in dimension starting from the primary by acting with $P$, and down in dimension acting with $K$.

We take the primary to be normalized to unity in the inner product. Denoting the conjugate $| \Delta\ra^{a_1\cdots a_{2s}\dag}= {}_{a_1\cdots a_{2s}}\la  \Delta |$, this normalization condition reads
\be {}_{b_1\cdots b_{2s}}\la  \Delta | \Delta\ra^{a_1\cdots a_{2s}}=\delta^{(a_1}_{\ b1}\cdots\delta^{a_{2s)}}_{\ b_{2s}} \, .\ee
The inner product of any of the descendants can then be computed by using the conjugation properties \eqref{PDJcong} and the conformal algebra commutators.  At certain values of $\Delta$, some of the descendent states become null.  These null states can then be factored out and the result is a shortened multiplet.  The values of $\Delta$ at which this happens are catalogued in \cite{Penedones:2015aga}.   Of interest to us are the multiply conserved currents \eqref{bosonmultiplyconsce}, \eqref{fermionmultiplyconsce} dual to partially massless fields of depth $t$, which have null states at level $s-t$.  We denote the conformal representation built from the primary\footnote{Unitary representations of the conformal group are all of this type \cite{Simmons-Duffin:2016gjk}, however there exist non-unitary representations which cannot be built by the action of $P$'s on some primary.  They contain states which are neither primary nor descendent.  An example is the so-called extended modules present in higher order free CFTs, which can be thought of as a kind of gluing together of ordinary modules \cite{Brust:2016gjy,Basile:2018eac}.  The AdS dual is a field which cannot be diagonalized \cite{Brust:2016zns}.} $ | \Delta\ra^{a_1\cdots a_{2s}}$ as $[s]_\Delta$.

\subsection{Generic superconformal representations}

The superconformal algebra is larger than the conformal algebra, and so the superconformal representations are larger than the conformal representations and can be built by joining together the conformal representations using the action of the generators $Q$ and $S$ which are not present in the conformal algebra.   The superconformal representation is built by starting with a superconformal primary, which is a conformal primary which in addition is annihilated by $S$,
\be S_a|\Delta\ra^{a_1\cdots a_{2s}}=0.\ee
We get the rest of the conformal primaries present in the superconformal multiplet by acting with $Q$'s in all possible combinations:
\be Q^a|\Delta\ra^{a_1\cdots a_{2s}},\ \ \ Q^{[a} Q^{b]}|\Delta\ra^{a_1\cdots a_{2s}}+(P|\Delta\ra\ {\rm terms}).\label{allqactiononpeq}\ee
The $P|\Delta\ra$ terms in \eqref{allqactiononpeq} are necessary and fixed by the requirement that the state be a conformal primary (i.e. annihilated by $K$), and we will be explicit about these terms shortly.
Each time we act by $Q$, $\Delta$ increases by $1/2$.  We need only consider the action by anti-symmetric combinations of the $Q$'s because the symmetric ones give descendants, due to the $\{Q,Q\}\sim P$ anti-commutation relation in \eqref{mainsssyq}.  We do not have to act with more than two $Q$'s because we cannot anti-symmetrize three or more $Q$'s.  

The states $Q^a|\Delta\ra^{a_1\cdots a_{2s}}$ with weight $\Delta+{1\over 2}$ at the first level break up into two irreducible representations, of spin $s\pm {1\over 2}$,
\be Q_a|\Delta\ra^{aa_2\cdots a_{2s}},\ \ \ Q^{(a}|\Delta\ra^{a_1\cdots a_{2s})}\,.\label{firsttwolevel1desce}\ee
These are conformal primaries which are not superconformal primaries, i.e. they are annihilated by $K$ but not by $S$.

Using \eqref{antisymsqeide} the states at the second level are a spin $s$ state with weight $\Delta+1$, which we call $|\Delta\ra_1^{a_1\cdots a_{2s}}$,
\be |\Delta\ra_1^{a_1\cdots a_{2s}}\equiv Q^aQ_a|\Delta\ra^{a_1\cdots a_{2s}},\label{level2descq2ge}\ee
as well as the first $P$ descendent of $|\Delta\ra^{a_1\cdots a_{2s}}$ which has the same quantum numbers as $|\Delta\ra_1^{a_1\cdots a_{2s}}$, which we call $|\Delta\ra_2^{a_1\cdots a_{2s}}$,
\bea   |\Delta\ra_2^{a_1\cdots a_{2s}} && \equiv  2s P^i  \sigma_{i\ b}^{(a_1}|\Delta\ra^{a_2\cdots a_{2s})b} \nn\\
&&=P^i  \sigma_{i\ b}^{a_1}|\Delta\ra^{b a_2\cdots a_{2s}}+P^i  \sigma_{i\ b}^{a_2}|\Delta\ra^{a_1b a_3\cdots a_{2s}}+\cdots+P^i  \sigma_{i\ b}^{a_{2s}}|\Delta\ra^{b a_2\cdots a_{2s-1}b}\, . \label{delta2defeqn}
\eea
There is one linear combination of $|\Delta\ra_1^{a_1\cdots a_{2s}}$ and $|\Delta\ra_2^{a_1\cdots a_{2s}}$ which is annihilated by $K$, and which is the true second level conformal primary.  By acting with $K$ and demanding the result vanish, we find this conformal primary, which we call $ |\tilde \Delta\ra^{a_1\cdots a_{2s}}$,
\be  |\tilde \Delta\ra^{a_1\cdots a_{2s}}= |\Delta\ra_1^{a_1\cdots a_{2s}}+{1\over \Delta-1} |\Delta\ra_2^{a_1\cdots a_{2s}}.\ee
(We will see later what happens when $\Delta=1$.)

Thus the generic superconformal representation contains four conformal primaries, and the structure of this representation can be illustrated as follows,
\be
\left\{ s\right\}_\Delta\ {\rm generic\ multiplet:}\ \ \ \ \ \ \ \ \raisebox{-96pt}{\epsfig{file=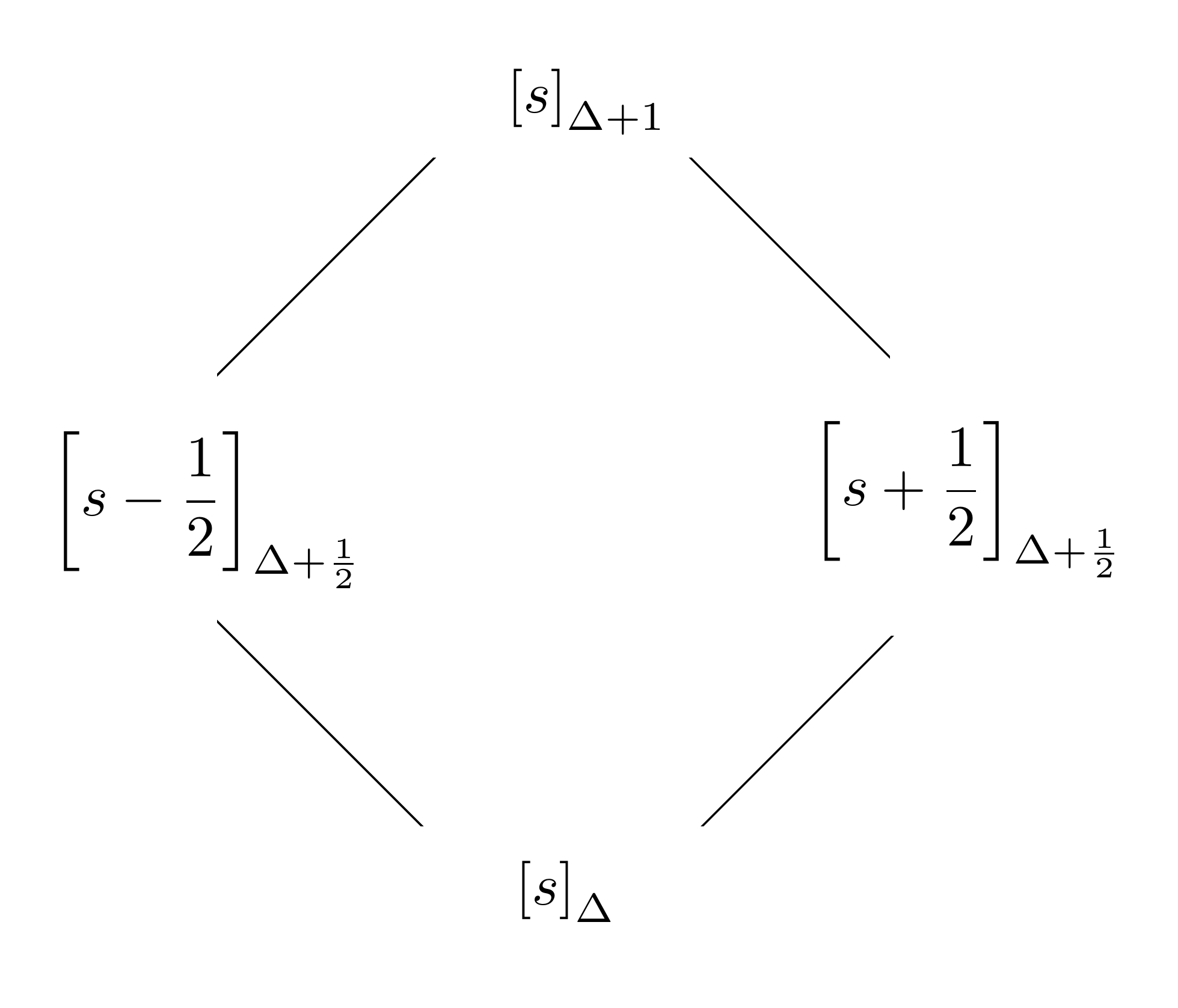,height=2.7in,width=3in}}\label{genericlongpice}
\ee
In this diagram the conformal dimension increases upward and the spin increases towards the right.  The superconformal primary is at the bottom.
We use the notation $\left\{ s\right\}_\Delta$ to denote the full superconformal multiplet, where $s,\Delta$ are the spin and dimension of the superconformal primary.  The notation $\left[ s\right]_\Delta$ is used for the individual conformal primaries within the superconformal multiplet.

In AdS$_4$, the multiplet \eqref{genericlongpice} corresponds to a generic multiplet of massive fields.  For example, the massive vector multiplet of AdS$_4$ corresponds to the case $\left\{{1\over 2}\right\}_{\Delta}$; in this multiplet there is a massive spin-1 $\left[1\right]_{\Delta+{1\over 2}}$ with 3 propagating degrees of freedom, a massive spin 0 $\left[0\right]_{\Delta+{1\over 2}}$ with one propagating degree of freedom, and two massive Weyl (or Majorana) fermions $\left[{1\over 2}\right]_{\Delta}$, $\left[{1\over 2}\right]_{\Delta+{1}}$ of spin-{1/2}, each with two propagating degrees of freedom, for a total of 4 bosonic and 4 fermionic propagating degrees of freedom.

In the case where the superconformal primary is spin 0, there is only one representation at the first level, $Q^a|\Delta\ra$, so there are only three conformal primaries in the representation, and the structure of the representation can be illustrated as follows
\be
\left\{ 0\right\}_\Delta\ {\rm generic\ scalar\ multiplet:}\ \ \ \ \ \ \ \ \raisebox{-90pt}{\epsfig{file=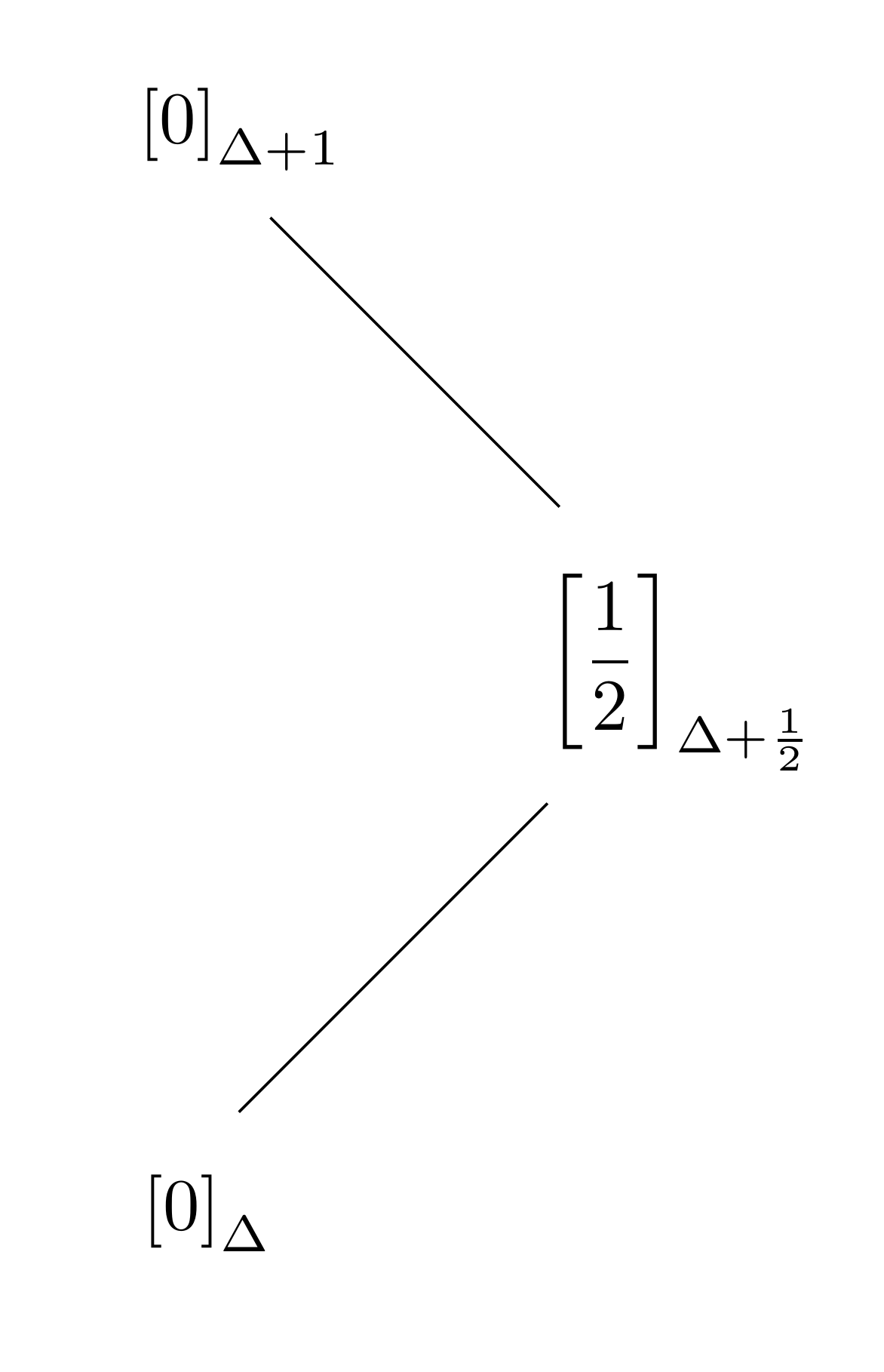,height=2.5in,width=1.7in}}
\ee
The conformal primary $\left[0\right]_{\Delta+1}$ at the second level is just $Q^a Q_a|\Delta\ra$.  In AdS$_4$, this corresponds to the generic massive Wess-Zumino multiplet; there are two scalars (which on flat space would be of equal mass packaged into a complex scalar, but on AdS have different masses) and one massive fermion, for a total of 2 bosonic and 2 fermionic propagating degrees of freedom.

\subsection{Shortening conditions}

Assuming that the superconformal primary is normalized to unity, the norms of the remaining conformal primaries within the superconformal multiplet can be computed using the conjugation properties \eqref{PDJcong} and the commutators of the superconformal algebra.  When the dimension $\Delta$ of the superconformal primary takes on certain values, some of the other conformal primaries may have zero norm.  When this happens, the multiplet may shorten.  In this section, we classify the various cases for which the conformal primaries within the superconformal multiplet get zero norms.

\subsubsection{$s=0$}

We start with the case where the superconformal primary has spin $s=0$.
Calculating the norm of the first descendent  $Q^a|\Delta\ra$, we find
\be \left|Q^a |\Delta\ra\right|^2=
\la \Delta |S_b Q^a|\Delta\ra=\la \Delta |\left\{S_b, Q^a\right\}|\Delta\ra=2\delta^a_{\ b}\la \Delta |D|\Delta\ra =2\delta^{a}_{\ b}\Delta \la \Delta |\Delta\ra=2\delta^{a}_{\ b}\Delta\, .\label{norms01equation}\ee
We see that when $\Delta=0$ this is a zero norm state, and in this case the second descendent $Q^a Q_a|\Delta\ra$ also has zero norm.  Both of these zero norm states have zero inner product with the remaining states and are thus null states.  They form their own sub-module which can be factored out.  We denote this by,
\be
\left\{ 0\right\}_0\ {\rm vacuum\ multiplet:}\ \ \ \ \ \ \ \ \raisebox{-90pt}{\epsfig{file=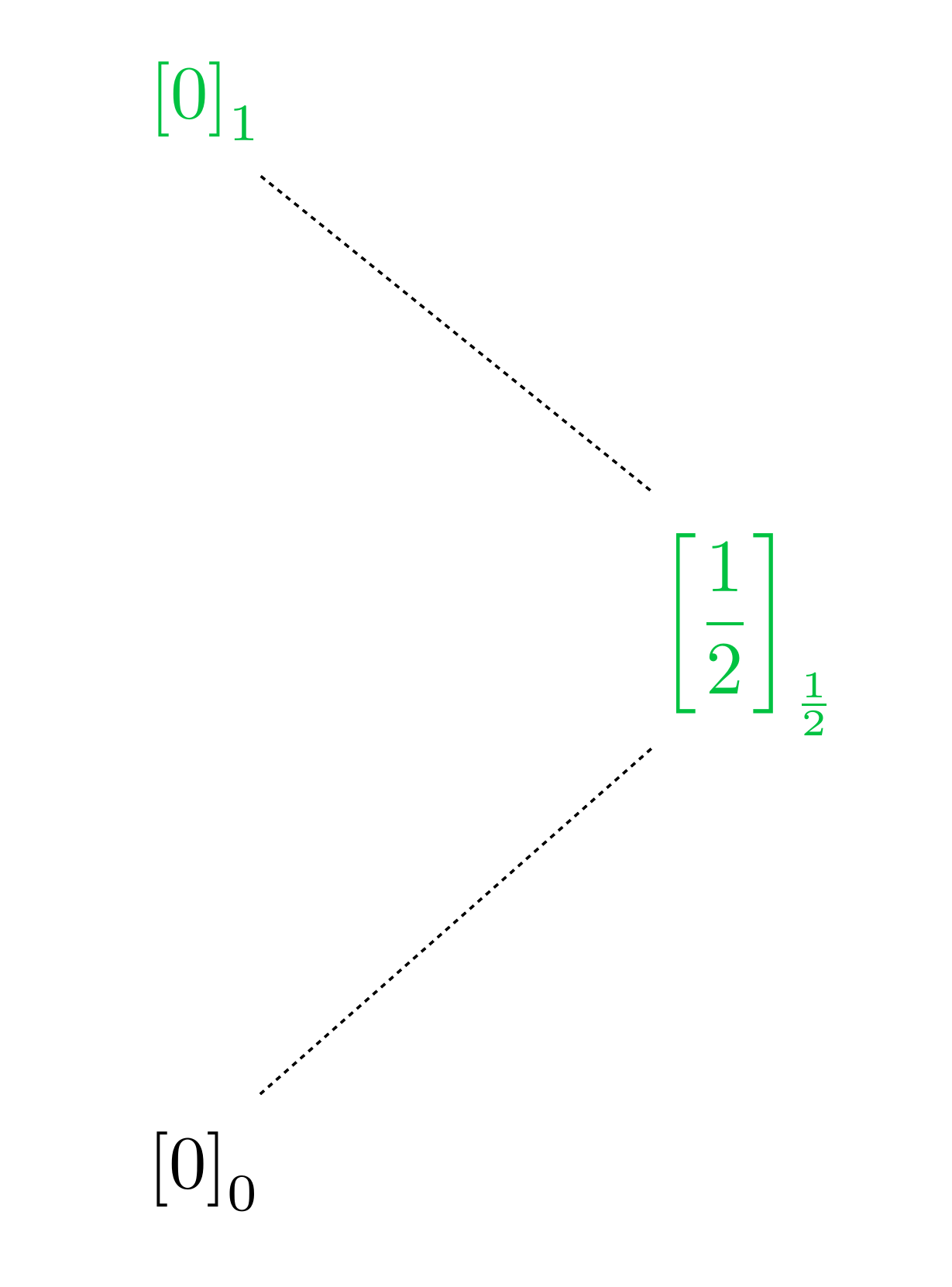,height=2.5in,width=1.7in}}\label{vacuummultiplete}
\ee
Here the null states are indicated in green, and they are removed from the multiplet.  Once they are removed we have a shortened multiplet consisting of only the state $\left[0\right]_0$.  A scalar conformal primary with $\Delta=0$ has no non-null conformal descendants, and so there is only one state in the entire supermultiplet \eqref{vacuummultiplete}.  This is the trivial vacuum multiplet.

Looking now at the second level, there is another shortening condition which we can find by looking at the norm of the second level descendant $Q^aQ_a|\Delta\ra$.  For this it is useful to first calculate an expression for $S^2Q^2$ between superconformal primaries (SCPs),
\be S^2Q^2\equiv S^bS_b Q^aQ_a \underset{\rm between\ SCPs}{=} -2\left\{Q^a,S^b\right\} \left\{Q_a,S_b\right\}+\left\{ Q^a,\left[\left\{Q_a,S_b\right\},S^b\right]\right\}\, .\ee
This can now be evaluated using the (anti)commutation relations of the superconformal algebra,
\be S^2Q^2 \underset{\rm between\ SCPs}{=} -16D^2+8J_{ij}J^{ij}+8D=-16\left[\Delta\left(\Delta-{1\over 2}\right)-s(s+1)\right],\ee
where we have used \eqref{Devprimconde} and \eqref{J2evprimconde} which say that $D=\Delta$ and $J_{ij}J^{ij}=2s(s+1)$ when acting on conformal primaries.  Thus, between scalar superconformal primaries where $s=0$, we have
\be \left|Q^2|\Delta\ra\right|^2=\la \Delta|\left(-S^b S_b \right)Q^a Q_a|\Delta\ra =16\Delta\left(\Delta-{1\over 2}\right).\label{norms02equation}\ee
We see again the vacuum multiplet shortening condition at $\Delta=0$, but now there is another root at $\Delta={1\over 2}$.  This is a shortening condition where the second level descendent goes null without the first one going null, and we have the shortened multiplet
\be
\left\{ 0\right\}_{1\over 2}\ {\rm singleton\ multiplet:}\ \ \ \ \ \ \ \ \raisebox{-90pt}{\epsfig{file=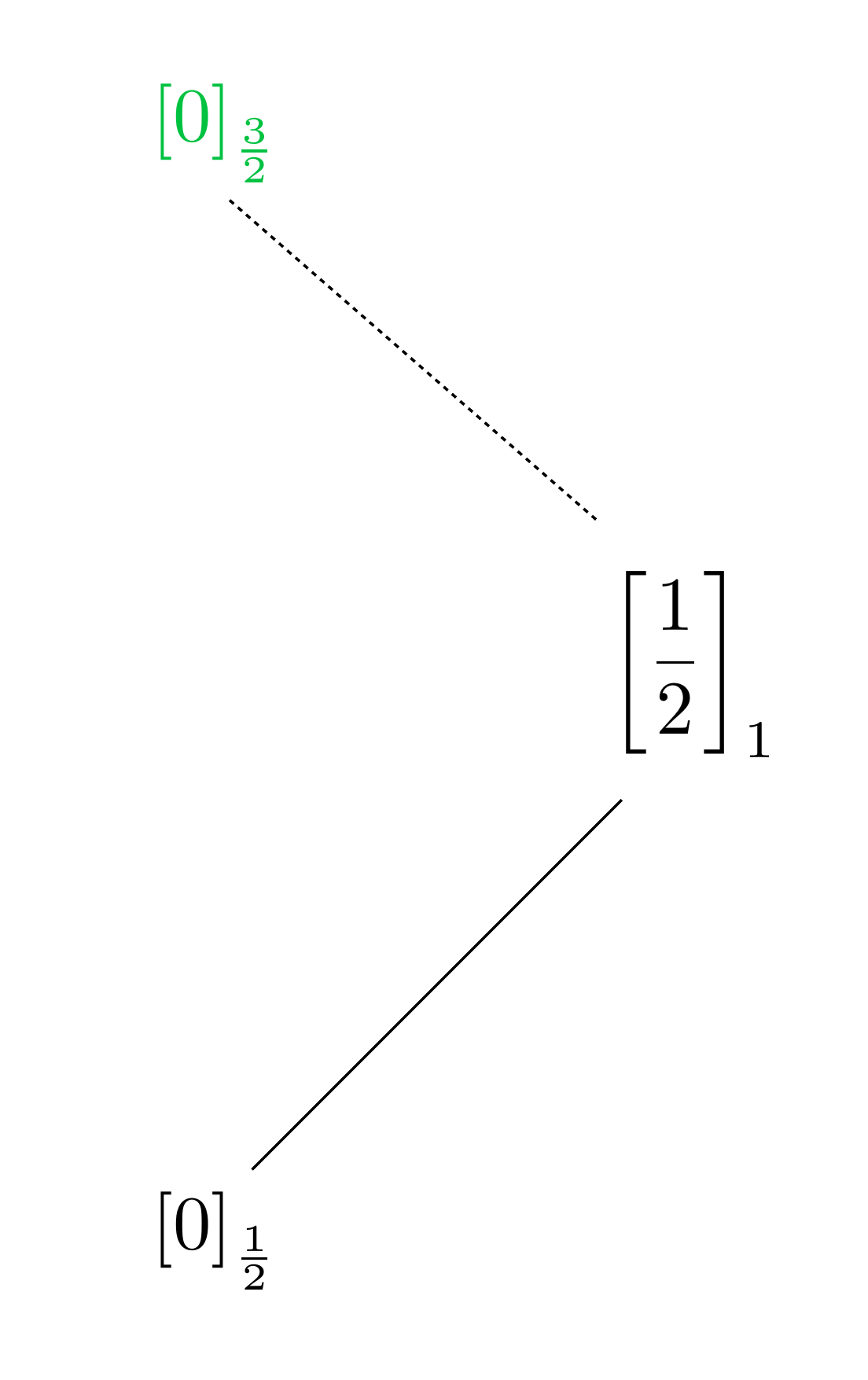,height=2.5in,width=1.6in}}
\ee
This is the supersymmetric Dirac singleton \cite{Dirac:1963ta}.

\begin{figure}[h!]
\begin{center}
\epsfig{file=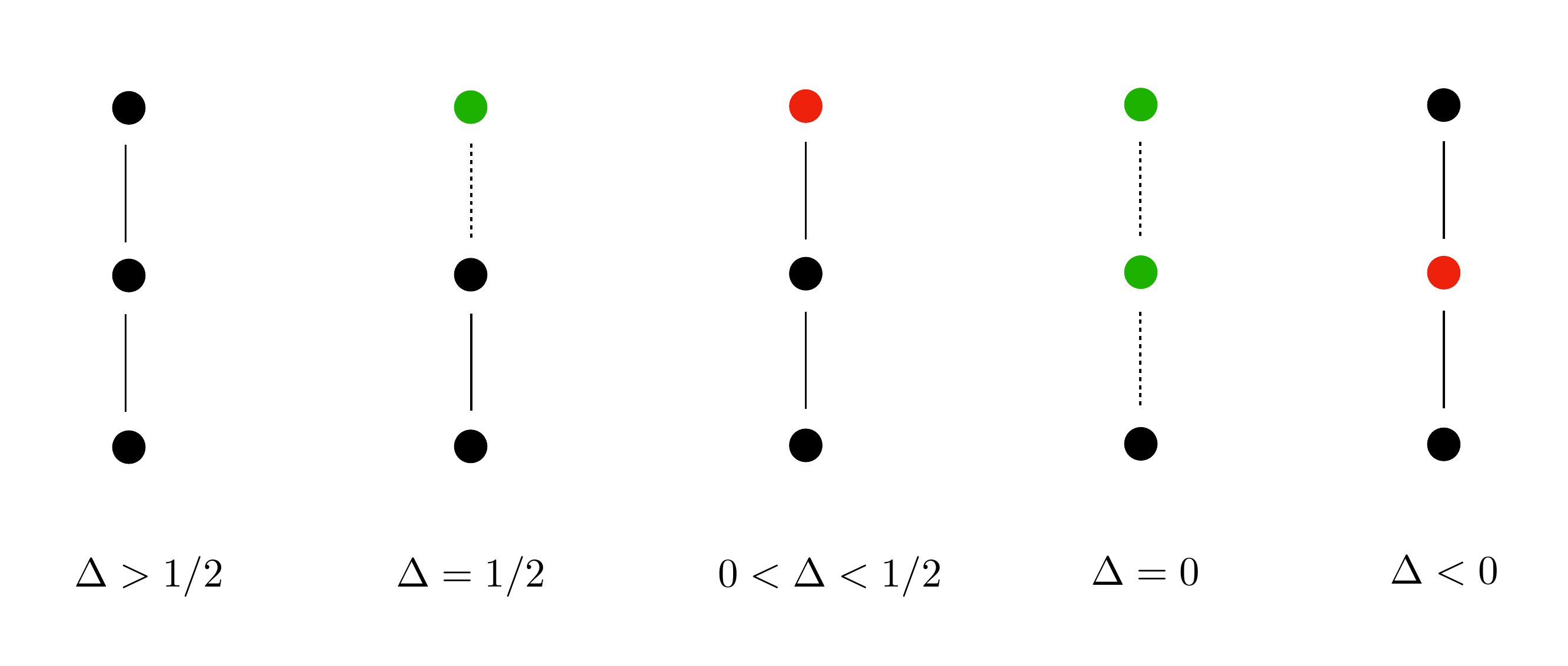,height=2.0in,width=5.0in}
\caption{\small Relative norms of the conformal primaries within the $\left\{0\right\}_\Delta$ multiplets, for the various values of $\Delta$.  Black is positive norm, red is negative norm, and green is zero norm.  We see the unitarity bound $\Delta\geq 1/2$.  All the cases below this bound have a negative norm conformal primary and are non-unitary, except for the vacuum multiplet at $\Delta=0$.}  
\label{norms0}
\end{center}
\end{figure}

As $\Delta$ crosses the values where zero norm states appear, the norms of various superconformal descendants will switch sign.  From \eqref{norms01equation} and \eqref{norms02equation}, we can read off the relative signs of the kinetic energy terms expected in the AdS$_4$ duals.  We see that when $\Delta>1/2$, all the norms are positive, consistent with unitarity.  When $\Delta=1/2$ we reach the first shortening condition where the second level superconformal descendant $\left[0\right]_{\Delta+1}$ goes null, and the other state $\left[{1\over 2}\right]_{\Delta+{1\over 2}}$ retains positive norm, so this multiplet is unitary.
When $0<\Delta<1/2$, the norm of $\left[{1\over 2}\right]_{\Delta+{1\over 2}}$ is still positive, while that of $\left[0\right]_{\Delta+{1}}$ becomes negative.  At $\Delta=0$ we hit the second shortening condition where we have the unitary vacuum multiplet.  When $\Delta<0$, the state $\left[{1\over 2}\right]_{\Delta+{1\over 2}}$ goes to negative norm, whereas $\left[0\right]_{\Delta+{1}}$ turns back to positive.  These norms are illustrated in figure \ref{norms0}.

\subsubsection{$s\geq 1/2$}

We now turn to the case where the superconformal primary has spin $s\geq 1/2$.  Starting at the first level, there are two possible shortening conditions, corresponding to when one or the other of the two first level descendants in \eqref{firsttwolevel1desce} go null.  
For the norm of $Q_a|\Delta\ra^{aa_1\cdots a_{2s-1}}$, we find
\bea \left| Q_a|\Delta\ra^{aa_1\cdots a_{2s-1}}\right|^2 &=&  {}_{bb_1\cdots b_{2s-1}}  \la \Delta | (-S^b) Q_a|\Delta\ra^{aa_1\cdots a_{2s-1}}= -\, {}_{bb_1\cdots b_{2s-1}}  \la \Delta | \left\{ S^b, Q_a\right\}|\Delta\ra^{aa_1\cdots a_{2s-1}} \nn\\ 
&=& {2s+1\over s}\left[\Delta-(s+1)\right]\delta^{(a_1}_{\ b_1}\cdots \delta^{a_{2s-1})}_{\ b_{2s-1}} \, .\label{eq:norm1}
\eea
For the norm of $Q^{(a}|\Delta\ra^{a_1\cdots a_{2s})}$ we find
\bea \left|Q^{(a}|\Delta\ra^{a_1\cdots a_{2s})}\right|^2 &=&   {}_{(b_1\cdots b_{2s}} \la\Delta| S_{b)} Q^{(a}|\Delta\ra^{a_1\cdots a_{2s})}={}_{(b_1\cdots b_{2s}} \la\Delta|\left\{ S_{b)} ,Q^{(a}\right\}|\Delta\ra^{a_1\cdots a_{2s})} \nn\\  
&=&  2(\Delta+s) \delta^{(a}_{\ b}\delta^{a_1}_{\ b_1}\cdots \delta^{a_{2s})}_{\ b_{2s}}\, . \label{eq:norm2}
\eea

At the second level, it is useful to compute the Graham matrix between the states $|\Delta\ra_1^{a_1\cdots a_{2s}}$ and $|\Delta\ra_2^{a_1\cdots a_{2s}}$ defined in \eqref{level2descq2ge} and \eqref{delta2defeqn},
\bea  && {}_{b_1\cdots b_{2s}}\left(\begin{array}{cc}   {}_1\la \Delta|\Delta\ra_1 &   {}_1\la \Delta|\Delta\ra_2  \\   {}_2\la \Delta|\Delta\ra_1 &   {}_2\la \Delta|\Delta\ra_2 \end{array}\right)^{a_1\cdots a_{2s}} \nn\\ && =\left(\begin{array}{cc}  16\left[\Delta\left(\Delta-{1\over 2}\right)-s(s+1)\right]  &  -8s(s+1) \\  -8s(s+1) & 8s(s+1)(\Delta-1) \end{array}\right) \delta^{(a_1}_{\ b1}\cdots\delta^{a_{2s)}}_{\ b_{2s}}\,. \label{grahammatrixdefce}
\eea
For $\Delta\not=1$, this hermitian form is diagonalized by the conformal primary $|\tilde \Delta\ra^{a_1\cdots a_{2s}}$ and the descendent $|\Delta\ra_2^{a_1\cdots a_{2s}}$,
\bea 
&& {}_{b_1\cdots b_{2s}}\la \tilde \Delta |\tilde \Delta\ra^{a_1\cdots a_{2s}}=\frac{8 (2 \Delta -1) (\Delta -s-1) (\Delta +s)}{\Delta -1} \delta^{(a_1}_{\ b1}\cdots\delta^{a_{2s)}}_{\ b_{2s}}\, , \label{null2levelde}\\
&& {}_{b_1\cdots b_{2s}\ 2}\la  \Delta | \Delta\ra_2^{a_1\cdots a_{2s}}=8 s (s+1)(\Delta -1)  \delta^{(a_1}_{\ b1}\cdots\delta^{a_{2s)}}_{\ b_{2s}} \, , \label{eq:norm4}\\
&& {}_{b_1\cdots b_{2s}}\la  \tilde\Delta | \Delta\ra_2^{a_1\cdots a_{2s}}=0 \,.
\eea

We see from \eqref{null2levelde} that the second level conformal primary becomes null when $\Delta=s+1$ and when $\Delta=-s$, which are the two values when the first level conformal primaries become null, as we can see from \eqref{eq:norm1}, \eqref{eq:norm2}.   There is thus a shortened multiplet when $\Delta=s+1$ with the following structure,
\be
\left\{ s\right\}_{s+1}\ {\rm massless\ short\ multiplet:}\ \ \ \ \ \ \raisebox{-86pt}{\epsfig{file=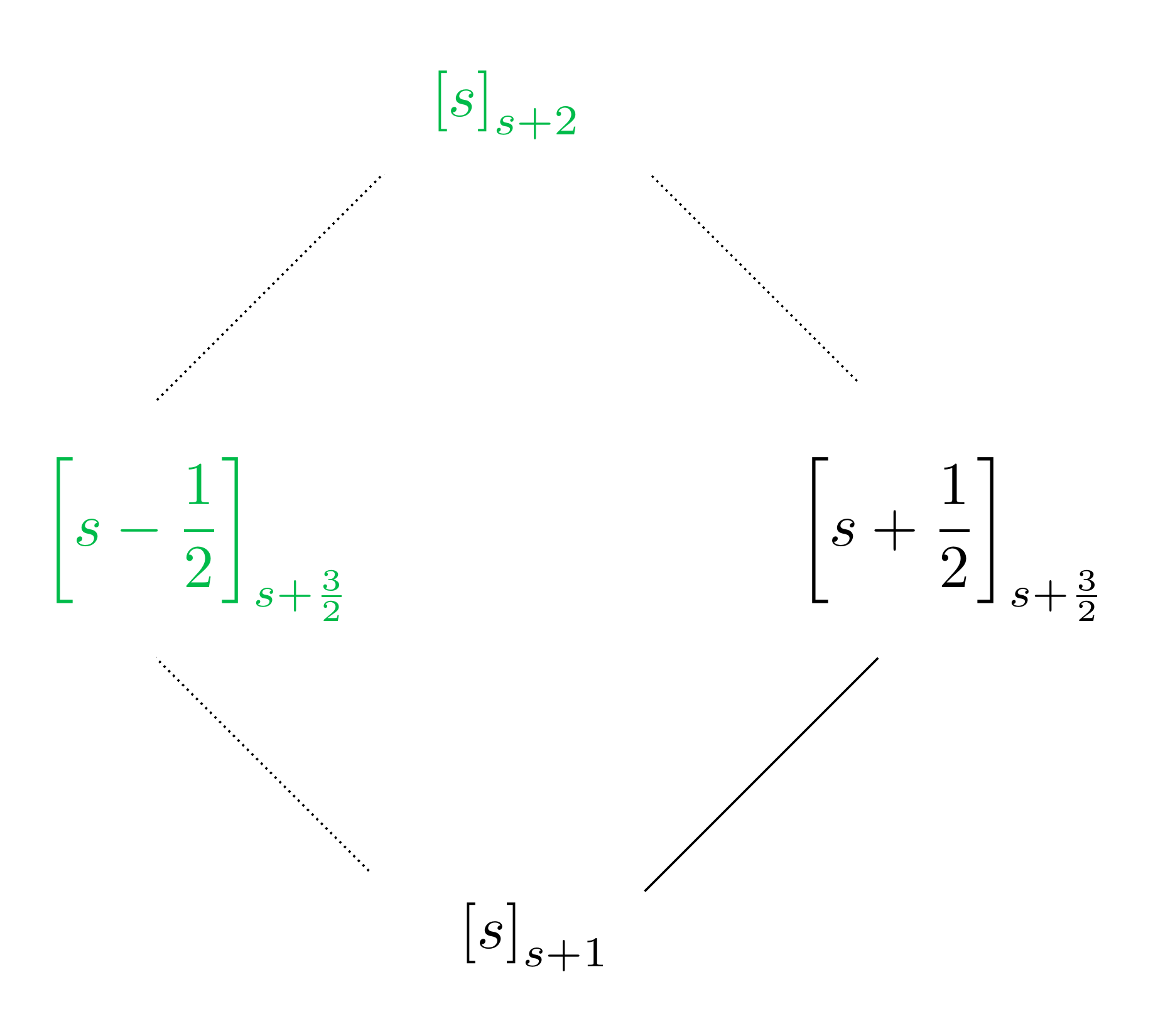,height=2.5in,width=2.5in}} \label{masslessmultipic2}
\ee
This is the conserved tensor multiplet, whose dual consists of massless fields in AdS$_4$.  For example, $\left\{ 3\over 2 \right\}_{5\over 2}$ is the ${\cal N}=1$ supergravity multiplet on AdS$_4$, consisting of a massless spin 3/2 and a massless graviton.

The shortened multiplet when $\Delta=-s$ has the following structure,
\be
\left\{ s\right\}_{-s}\ {\rm finite\ short\ multiplet:}\ \ \ \ \ \ \raisebox{-85pt}{\epsfig{file=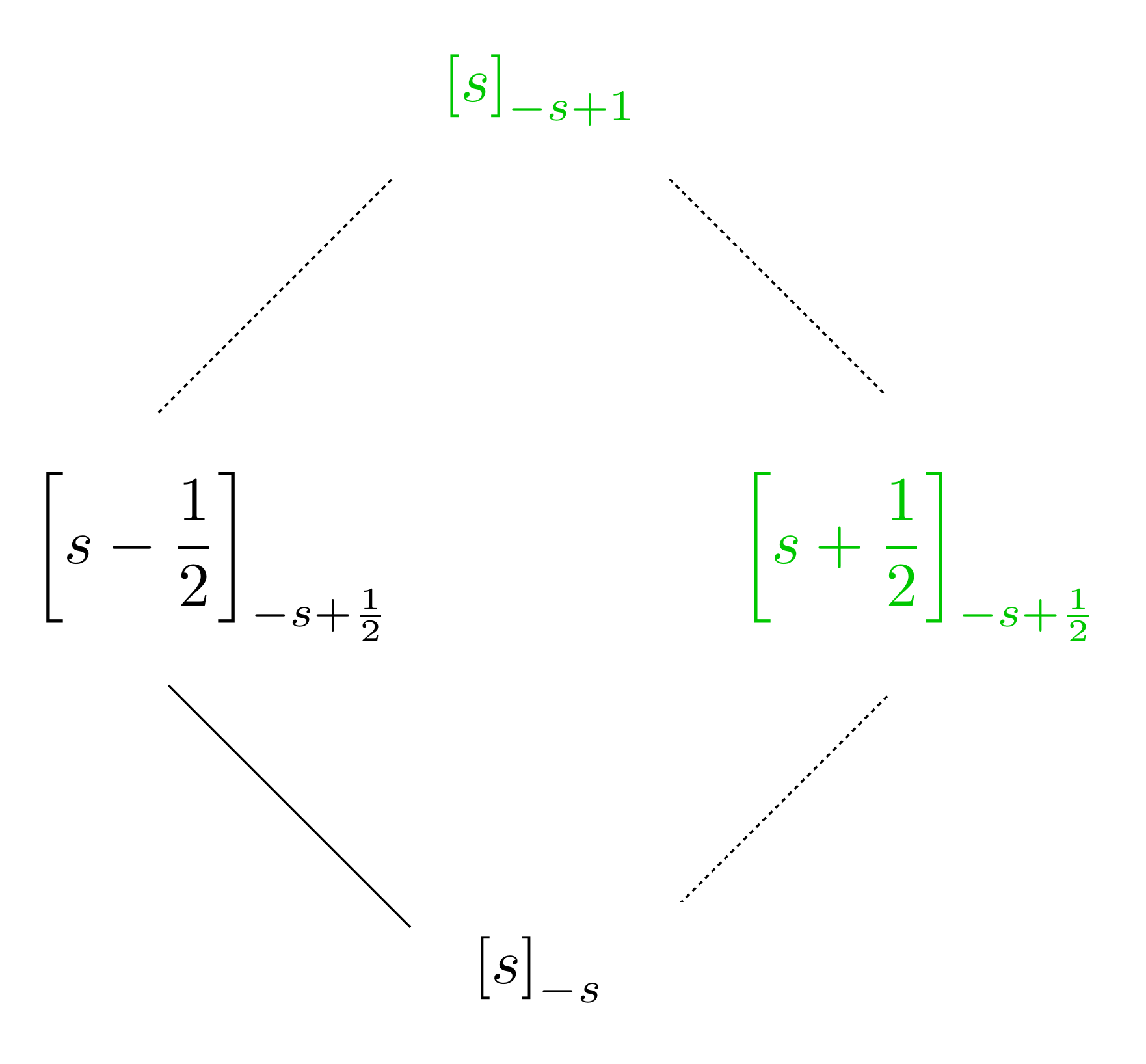,height=2.5in,width=2.7in}} \label{short2multe}
\ee
There are conformal multiplets which themselves have null descends which when factored out make the Verma module finite dimensional.  This is the type I shortening in the language of \cite{Penedones:2015aga}, which occurs for the conformal representations $\left[s\right]_{-s-k}$, $k=0,1,2,\cdots$.  These finite multiplets have dimension $\Delta<3/2$ and so occur for alternately quantized fields which have a finite number of modes on AdS \cite{Balasubramanian:1998sn,Brust:2016zns}.  The superconformal multiplets $\left\{s\right\}_{-s-1-k}$, $k=0,1,2,\cdots$ are long SUSY multiplets which link together these finite dimensional multiplets.  These are thus finite dimensional supermultiplets, which can be thought of as supersymmetric spherical harmonics on AdS$_4$.  The short multiplets \eqref{short2multe} are the boundary case, linking together two finite dimensional conformal multiplets on the boundary of the region of finite dimensional multiplets, thus they can be thought of as short supersymmetric spherical harmonics on AdS$_4$. 

We see also from \eqref{null2levelde} that the second level conformal primary becomes null when $\Delta=1/2$.  At this value, neither of the two first level conformal primaries are null, so we have the shortened multiplet:
\be
\left\{ s\right\}_{{1\over 2}}\ {\rm PM\ shadow\ short\ multiplet:}\ \ \ \ \ \raisebox{-100pt}{\epsfig{file=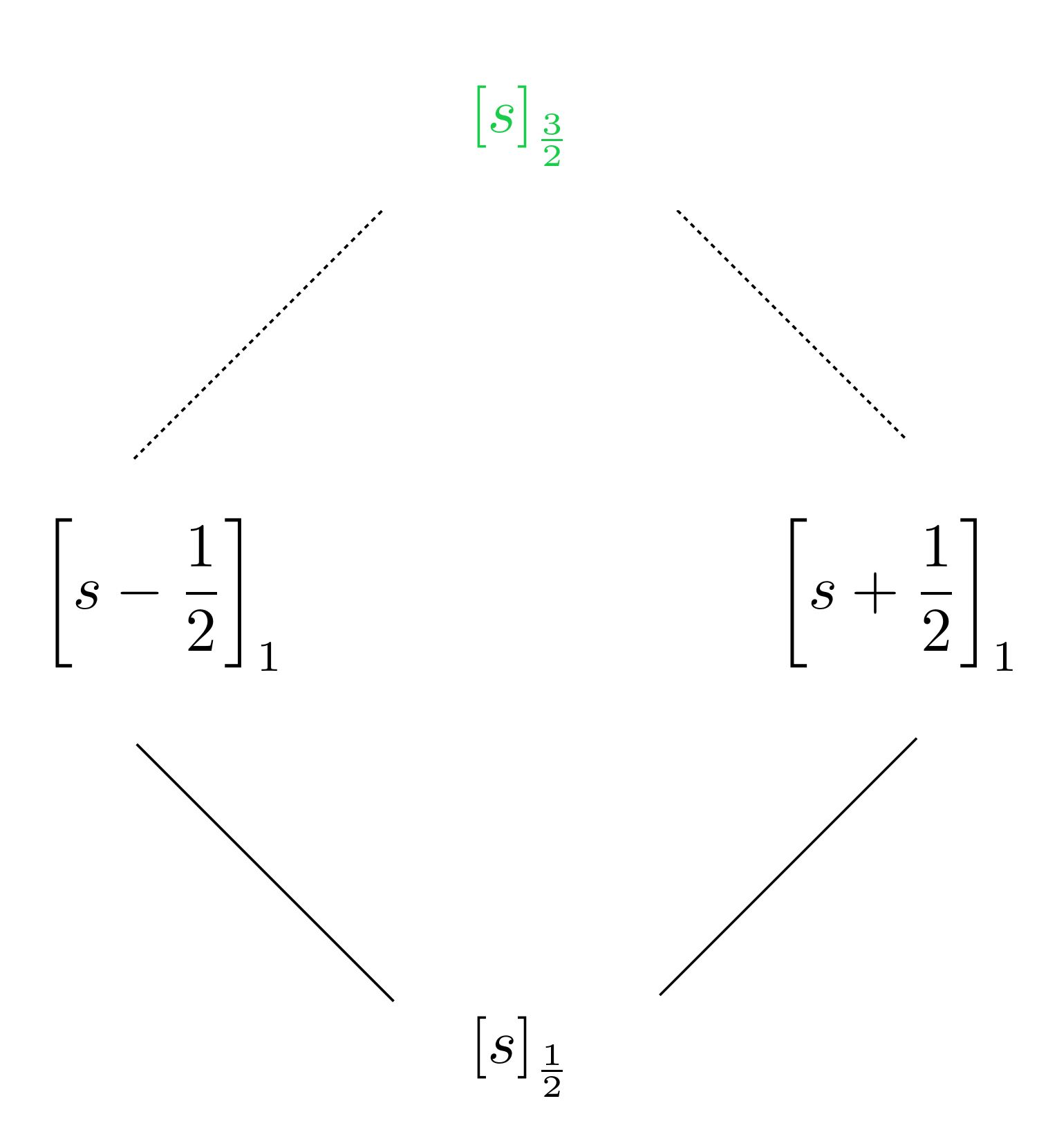,height=2.9in,width=2.9in}} \label{short3multe}
\ee
This multiplet also involves only $\Delta<3/2$ and so corresponds to alternately quantized fields.  In fact, the fields here are all alternately quantized partially massless fields of depth $t=0,1/2$ and so these are the shadow operators of multiply conserved currents.  These shadow operators themselves have shortening conditions corresponding to the vanishing of the partially massless field strengths \cite{Hinterbichler:2016fgl} (this is the type IV shortening in the Appendix of \cite{Penedones:2015aga}).  The short multiplet \eqref{short3multe} sits at the boundary of the region of multiplets of partially massless shadow operators.

Finally, we must treat the case $\Delta=1$.  In this case, there is a degeneration because the conformal primary $|\tilde \Delta\ra^{a_1\cdots a_{2s}}$ degenerates with the descendent $|\Delta\ra_2^{a_1\cdots a_{2s}}$, and this state has zero norm: ${}_{b_1\cdots b_{2s}\ 2}\la  \Delta | \Delta\ra_2^{a_1\cdots a_{2s}}=0$.  However, as we can tell from the Graham matrix \eqref{grahammatrixdefce}, this state is not null because it does not have zero inner product with the other states, and so it does not decouple.  In fact there is a second null state $|\Delta_N\ra^{a_1\cdots a_{2s}}$, linearly independent of $| \Delta\ra_2^{a_1\cdots a_{2s}}$, which is not orthogonal to $|\Delta\ra_2^{a_1\cdots a_{2s}}$, and is neither primary nor descendent,
\be |\Delta_N\ra^{a_1\cdots a_{2s}}\equiv | \Delta\ra_1^{a_1\cdots a_{2s}}-\frac{2 s(s+1)-1}{2 s (s+1)}| \Delta\ra_2^{a_1\cdots a_{2s}},\ee
\be  {}_{b_1\cdots b_{2s}}\la  \Delta_N | \Delta_N\ra^{a_1\cdots a_{2s}}=0,\ \ \ \ \ee
\be {}_{b_1\cdots b_{2s}}\la  \Delta_N | \Delta\ra_2^{a_1\cdots a_{2s}}=-8s(s+1) \, \delta^{(a_1}_{\ b1}\cdots\delta^{a_{2s)}}_{\ b_{2s}}\, .\ee
The Graham matrix is Lorentzian.  Diagonalizing it would produce a state with positive norm and a state with negative norm, though these would not have nice actions under the superconformal generators.  This phenomenon can only happen in the non-unitary case where some states have negative norms.  The structure of the module is that of an ``extended module'' rather than a shortening.  
This is a supersymmetric version of the extended modules found in the Hilbert space of higher order free CFTs \cite{Brust:2016gjy}.  Note that this case bridges the $\Delta=3/2$ divide between the alternately quantized fields and the standardly quantized fields.  

\begin{figure}[h!]
\begin{center}
\epsfig{file=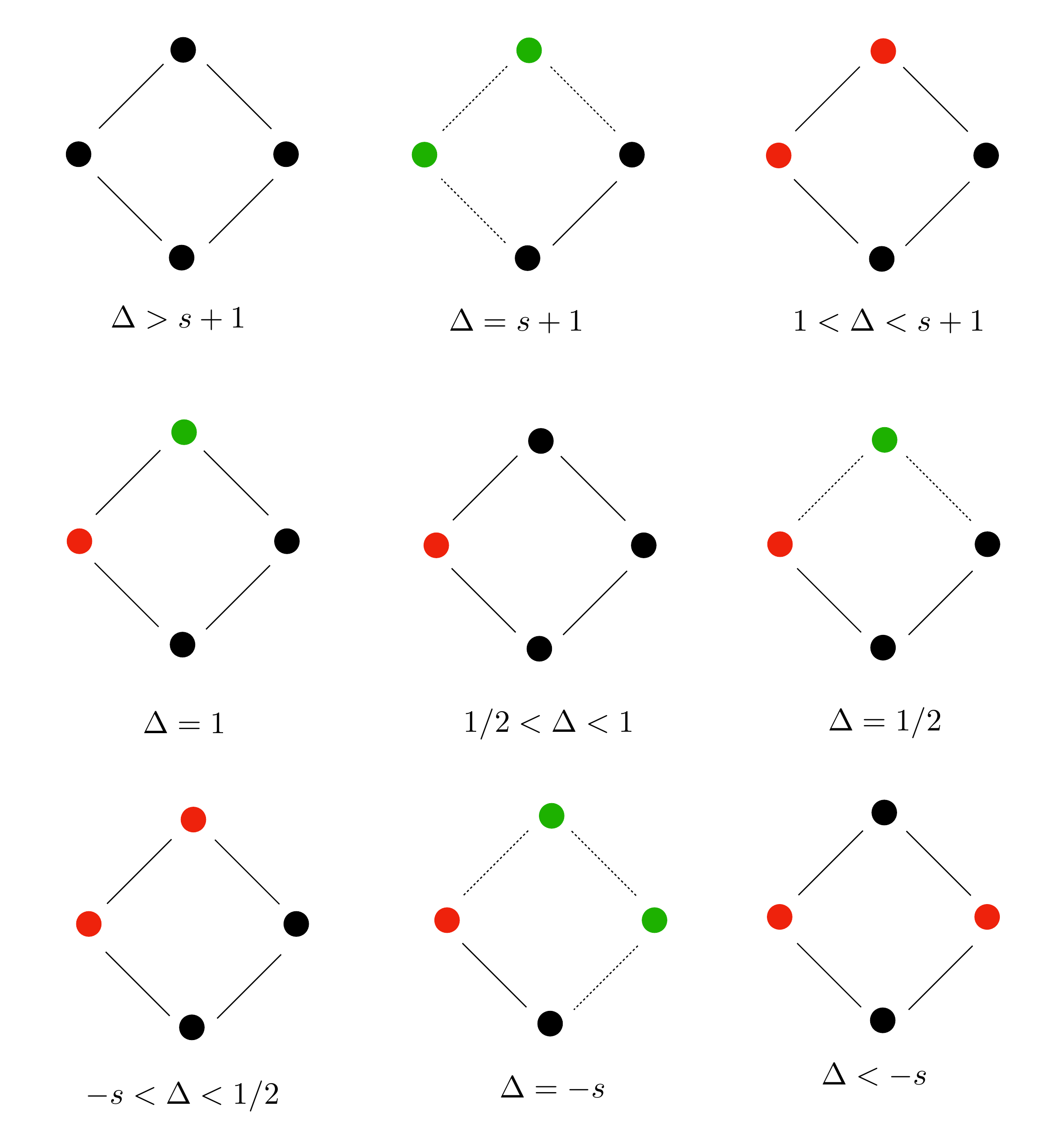,height=5.0in,width=4.8in}
\caption{\small Relative norms of the conformal primaries within the $\left\{s\right\}_\Delta$ multiplets with $s\geq 1/2$, for the various values of $\Delta$.  Black is positive norm, red is negative norm, and green is zero norm.  We see the unitarity bound $\Delta\geq s+1$; all the multiplets below this bound have at least one negative norm conformal primary and are thus non-unitary.}  
\label{normss}
\end{center}
\end{figure}

Finally, we can read off from \eqref{eq:norm1}, \eqref{eq:norm2}, \eqref{null2levelde} the relative norms for various values of $\Delta$, which correspond to the signs of the kinetic energy terms expected in AdS$_4$.  These norms are illustrated in Figure \ref{normss}.

\section{Partially Massless Multiplets}

With the representations in hand, we can now look for those that contain the partially massless fields.  In what follows, we will restrict to those multiplets which contain only $\Delta\geq 3/2$ states, so that we are dealing only with standardly quantized fields with a clean AdS$_4$ interpretation.

\begin{figure}[h!]
\begin{center}
\epsfig{file=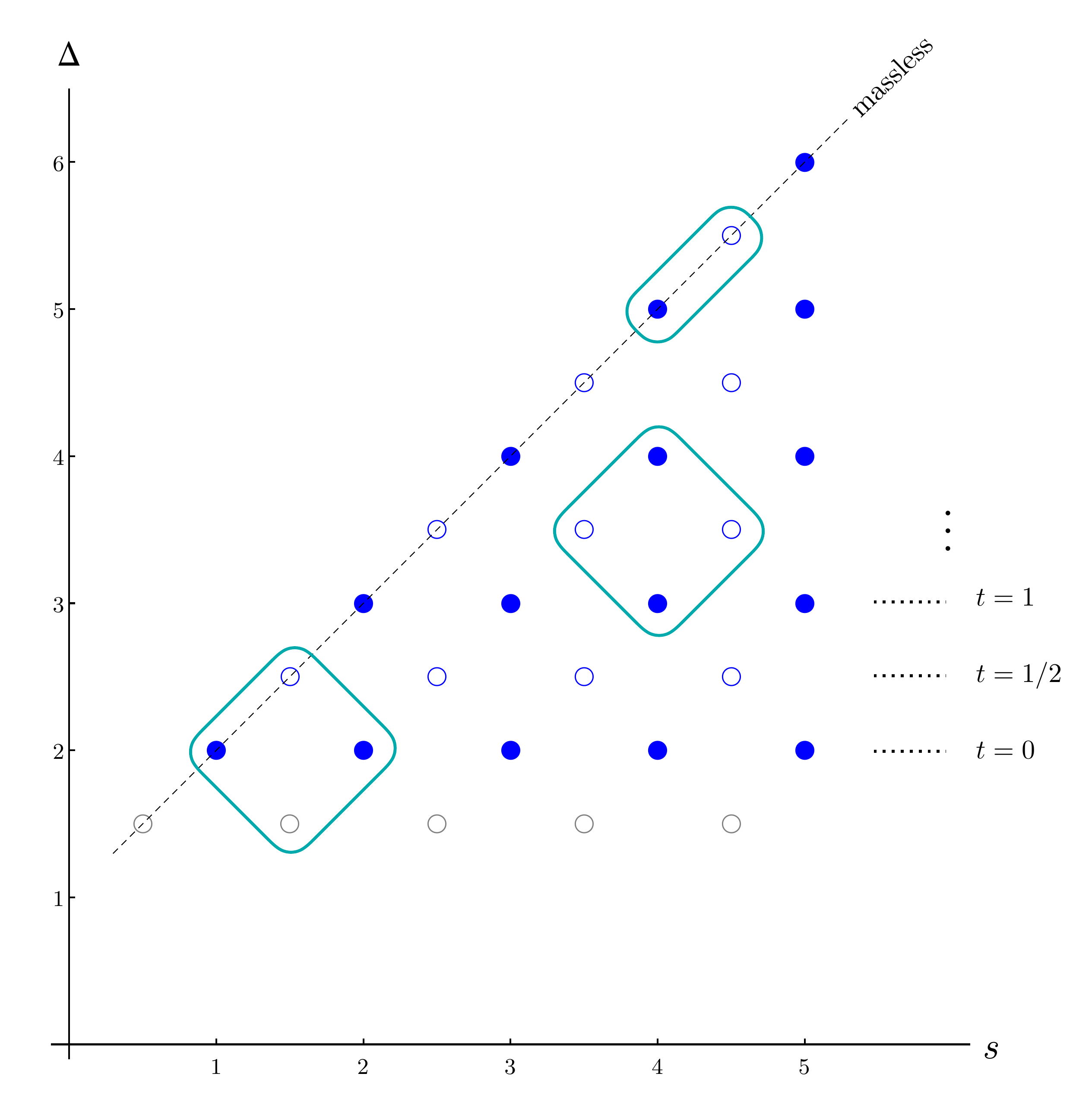,height=5.0in,width=5.0in}
\caption{\small The ${\cal N}=1$ partially massless multiplets.  The filled circles are bosons, the hollow circles are fermions.  The blue circles are the partially massless points, the grey open circles on the bottom row at $\Delta=3/2$ are massive fermions that participate in some of the multiplets.  The multiplets are all diamonds containing four fields, like the example diamonds illustrated which show the $\left\{ 3\over 2 \right\}_{3\over 2 }$ partially massless spin-2 representation \eqref{PMspin2multpice} as well as the $\left\{ 4 \right\}_{3}$ representation which contains only partially massless fields.  The diamonds degenerate at the top due to the shortening condition \eqref{masslessmultipic2}.  These are the familiar unitary massless multiplets, and in these short multiplets there are only two massless particles, like the $\left\{ 4 \right\}_{5}$ example shown.}  
\label{plot1}
\end{center}
\end{figure}

The partially massless multiplets are shown in Figure \ref{plot1}.  Other than the massless fields, which appear in the shortened multiplet \eqref{masslessmultipic2}, the multiplets with standard quantization containing partially massless fields are not short multiplets.  The PM fields all appear in the generic long multiplet \eqref{genericlongpice}.  
For example, the simplest multiplet containing a partially massless spin-2 is the following:
\be
\left\{ 3\over 2\right\}_{3\over 2}\ {\rm PM\ graviton\ multiplet:}\ \ \ \ \ \ \raisebox{-102pt}{\epsfig{file=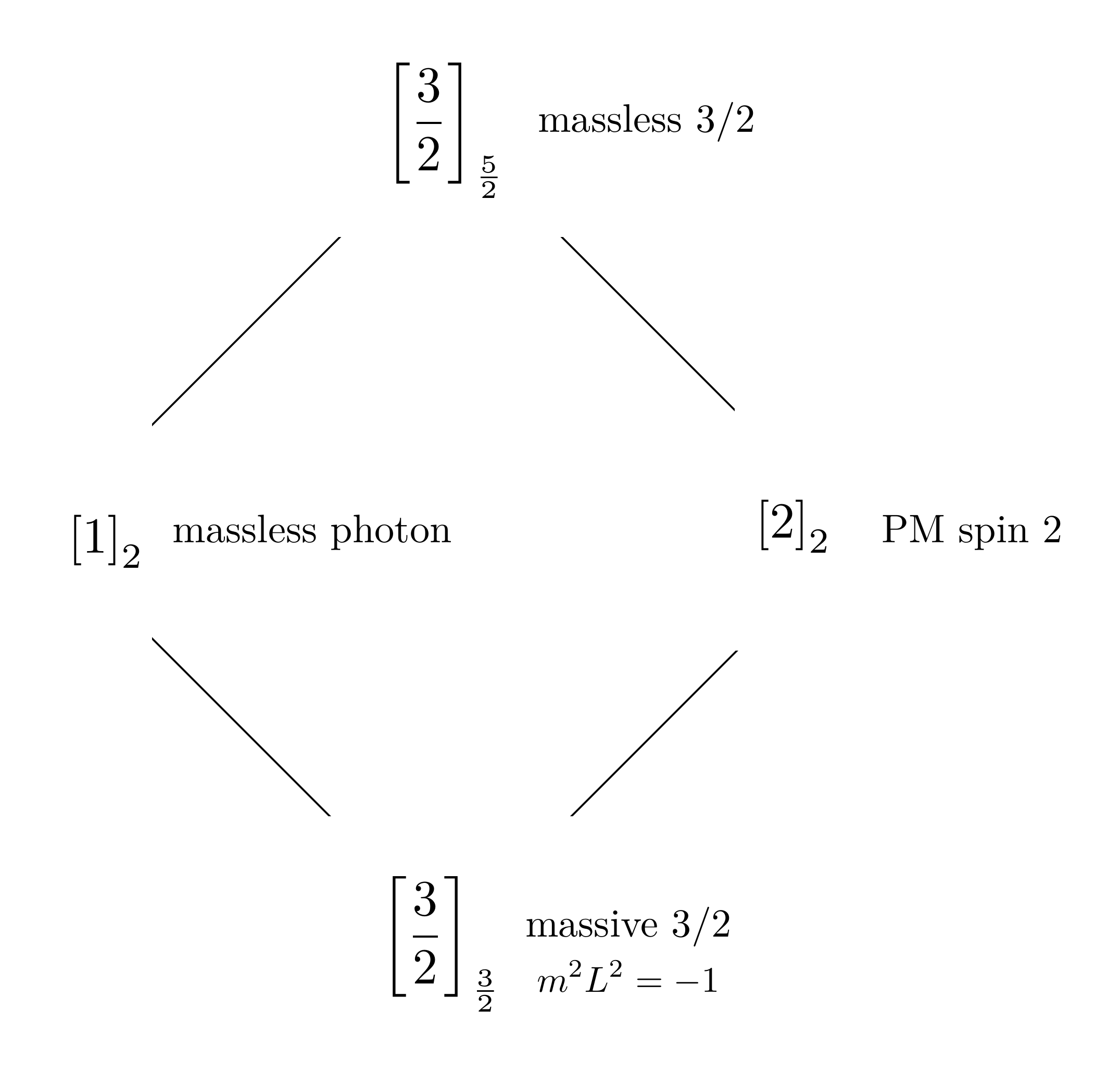,height=2.9in,width=3in}}\label{PMspin2multpice}
\ee
The propagating bosonic and fermionic degrees of freedom of the corresponding AdS$_4$ fields match: the PM spin-2 has 4 degrees of freedom and the photon has 2, whereas the massive spin-3/2 has 4 degrees of freedom and the massless spin-3/2 has 2.
The general pattern of PM field multiplets can be visualized from figure \ref{plot1}.  In all the multiplets, it can be checked explicitly that the number of bosonic and fermionic propagating degrees of freedom in AdS$_4$ match.  

The relative norms of the states within the partially massless multiplets can be read from Figure \ref{normss}.  The massless multiplets are unitary.  The partially massless multiplets we are considering all lie in the range $1<\Delta<s+1$, so as indicated in the figure, the states $\left[ s-{1\over 2}\right]_{\Delta+{1\over 2}}$ and $\left[ s\right]_{\Delta+{1}}$ have negative norm, whereas $\left[ s+{1\over 2}\right]_{\Delta+{1\over 2}}$ and the superconformal primary $\left[ s\right]_{\Delta}$ have positive norm.  Thus on AdS$_4$, the two bosonic fields within the multiplet will have opposite sign kinetic terms, and the two fermionic fields will have opposite signs.  We will see in Section \ref{PMadssection} how these opposite signs are necessary in order to be consistent with the reality properties of the fields and transformations on AdS$_4$.

\subsection{Branching rules\label{branchingsection}}

In the partially massless multiplets, the partially massless gauge parameters for all the fields themselves form a supermultiplet.  For example, for the partially massless spin-2 multiplet in \eqref{PMspin2multpice}, the gauge parameters form the following scalar multiplet:
\be
\left\{ 0 \right\}_{3 }\ {\rm PM\ graviton\ gauge\ parameter\ multiplet:}\ \ \ \ \ \ \raisebox{-102pt}{\epsfig{file=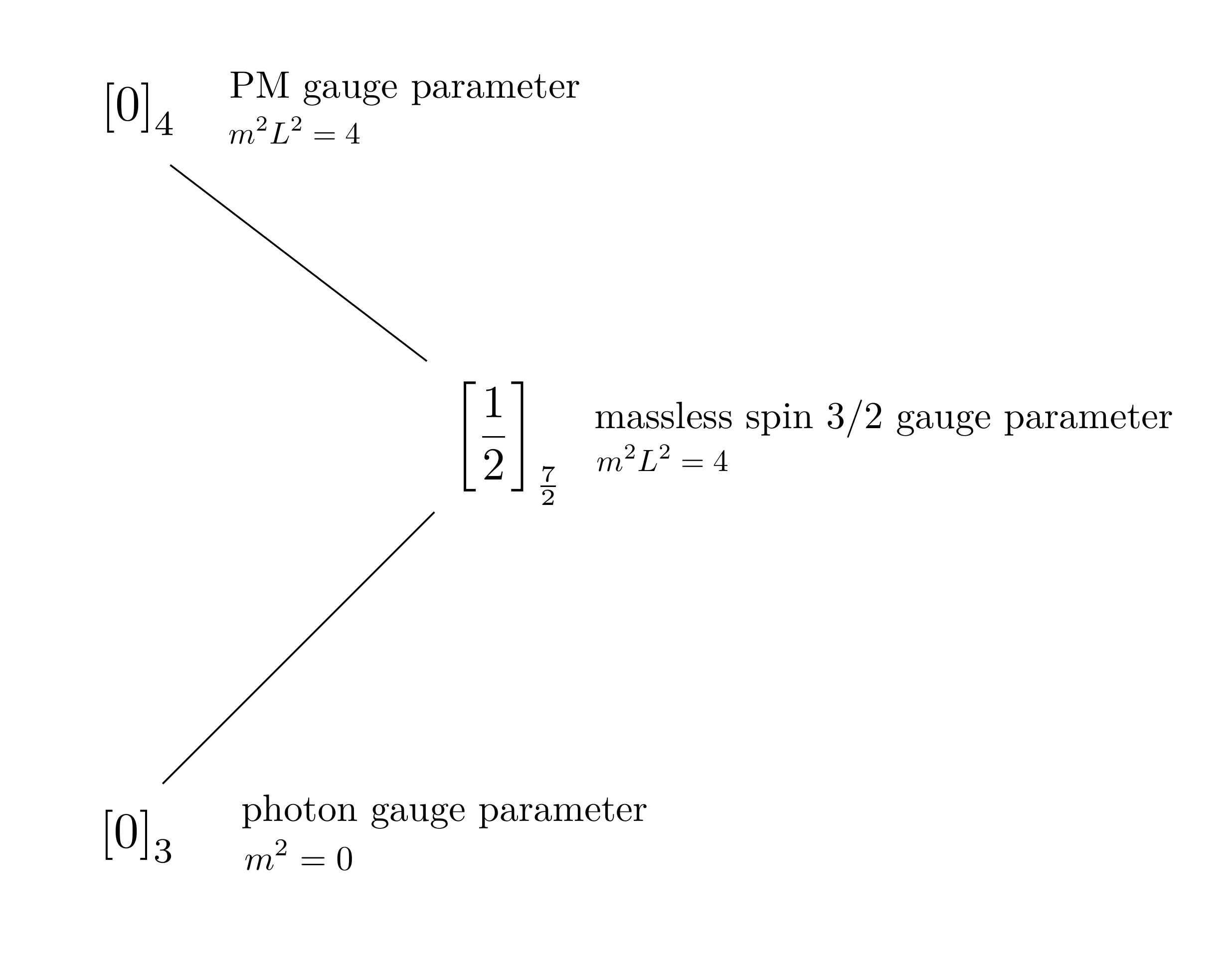,height=2.7in,width=3in}}\label{PMspin2multpicegauge}
\ee

As the dimension $\Delta$ of a generic massive graviton multiplet approaches the value $\Delta=3/2$ of the partially massless multiplet, conformal descendants of the spin-2, spin-1 and spin 3/2 conformal primaries become null.  On the AdS$_4$ side, the fields are becoming massless and partially massless, and so they are developing gauge symmetries.  The gauge parameters corresponding to the null descendent states can be thought of as \stu fields which are decoupling in the limit (see \cite{deRham:2018svs} for a recent discussion of this in context of the massless and partially massless spin 2). 

This describes a branching rule: as the generic massive multiplet approaches one of the partially massless values, it breaks up into the partially massless multiplet plus the multiplet corresponding to the emerging gauge parameters,
\be \left\{ 3\over 2 \right\}_{\Delta}\ \underset{\Delta \rightarrow {3\over 2}}{\rightarrow}\ \left\{ 3\over 2 \right\}_{3\over 2} \oplus \left\{ 0\right\}_{3}.\ee  
In terms of all the fields, the branching rule is as follows,
\be
\raisebox{-70pt}{\epsfig{file=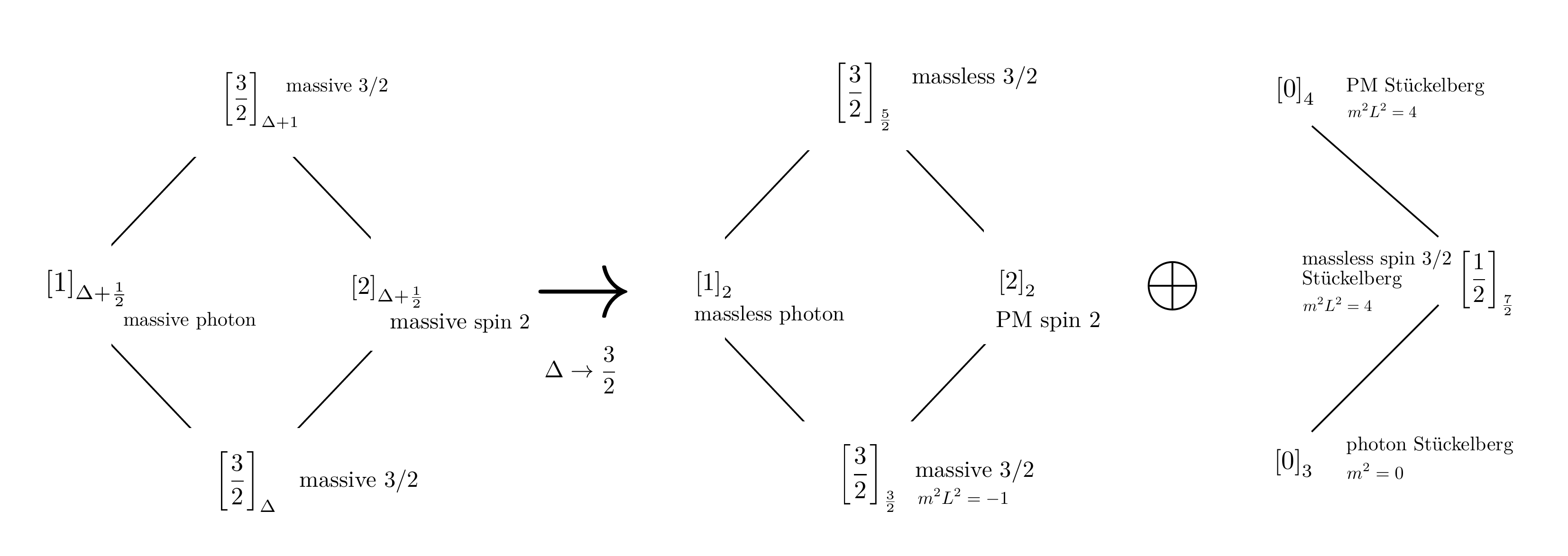,height=2.2in,width=6.3in}}\label{PMspin2multpicebranch1}
\ee
As the partially massless point $\Delta=3/2$ is approached, the massive $\left\{3\over 2\right\}_\Delta$ representation shortens into the $\left\{3\over 2\right\}_{3\over 2}$ representation, and the null states split off into the massive  $\left\{0\right\}_3$ representation.   Within this massive $\left\{0\right\}_3$ representation, the $\left[0\right]_4$ is the longitudinal mode of the $\left[2\right]_{\Delta+{1\over 2}}$ spin-2 which becomes partially massless, the $\left[0\right]_{3}$ is the longitudinal mode of the $\left[1\right]_{\Delta+{1\over 2}}$ spin-1 which becomes massless, and the $\left[1\over 2\right]_{7\over 2}$ is the longitudinal mode of the $\left[3\over 2\right]_{\Delta+1}$ spin-3/2 which becomes massless.

There is also a branching rule in the massless limit $\Delta \rightarrow {5/ 2}$, 
\be \left\{ 3\over 2 \right\}_{\Delta}\ \underset{\Delta \rightarrow {5\over 2}}{\rightarrow}\ \left\{ 3\over 2 \right\}_{5\over 2} \oplus \left\{ 1\right\}_{3}.\ee
In terms of the fields, this branching rule is as follows,
\be
\raisebox{-70pt}{\epsfig{file=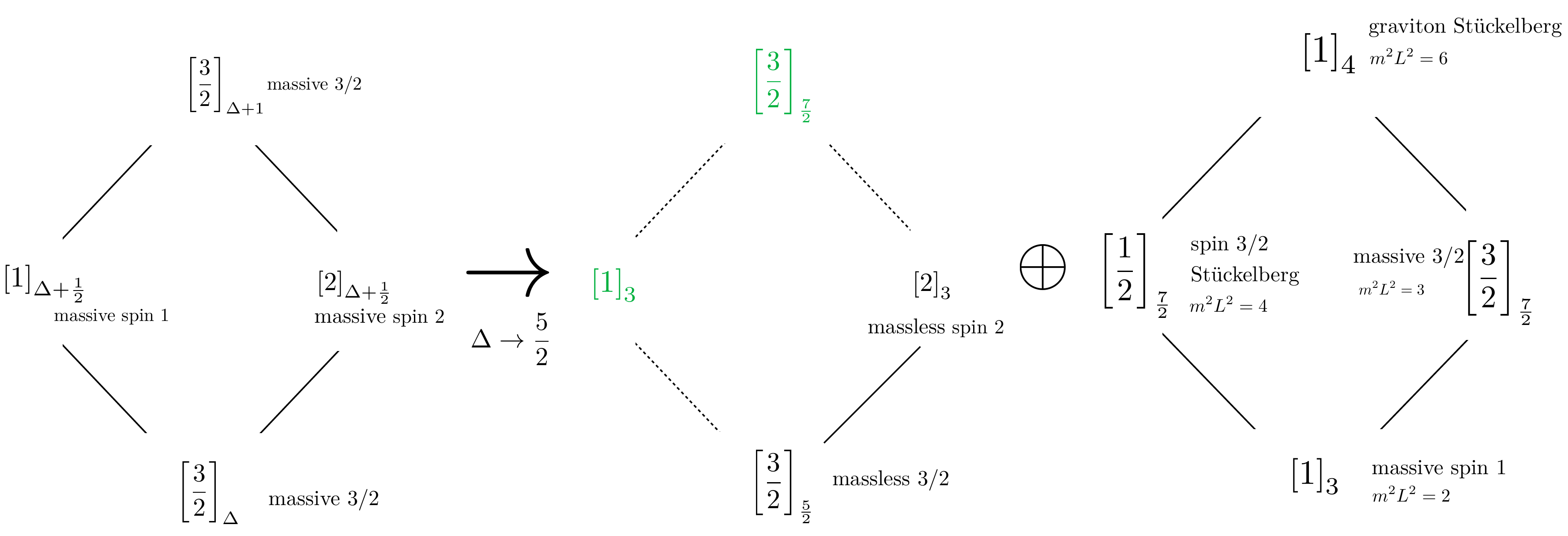,height=2.2in,width=6.5in}}\label{PMspin2multpicebranch2}
\ee
As the massless point $\Delta=5/2$ is approached, the massive $\left\{3\over 2\right\}_\Delta$ representation shortens into the $\left\{3\over 2\right\}_{5\over 2}$ representation, and the null states split off into the massive  $\left\{1\right\}_3$ representation.   Within this massive $\left\{1\right\}_3$ representation, the $\left[1\right]_4$ is the longitudinal mode of the massive spin-2 which becomes massless, the $\left[1\over 2\right]_{7\over 2}$ is the longitudinal mode of the massive spin-3/2 which becomes massless, and the $\left[1\right]_3$ and $\left[3\over 2\right]_{7\over 2}$ are the two conformal primaries of the $\left\{3\over 2\right\}_\Delta$ which go null.
In Section \ref{masslessadslimitsection}, we will see that in the AdS$_4$ action there will be fields which capture these null states, in addition to the \stu fields for the emerging gauge parameters, and all these fields together will split off into their own massive multiplet.

The general partially massless branching rule is
\be \left\{ s \right\}_{\Delta}\underset{\Delta \rightarrow {t+2}}{\rightarrow}\left\{ s \right\}_{t+2} \oplus \left\{ t+{1\over 2}\right\}_{s+{3\over 2}}\, .\ee
Here $s$, $t$ are the spin and depth of the PM point in figure \ref{plot1} corresponding to the superconformal primary (with $t=-{1\over 2}$ corresponding to the cases where the superconformal primary is one of the $\Delta=3/2$ fermions).
In the PM multiplets without massless fields, the gauge parameter multiplet contains precisely the gauge parameters for the partially massless fields.  The gauge parameter is always the reflection of the field about the line $\Delta=s+2$ in the $s,\Delta$ plane depicted in Figure \ref{plot1}, and so the gauge parameter multiplet is the reflection of the PM multiplet.  In the massless multiplets, the \stu multiplet contains, in addition to the gauge parameters for the two massless fields, \stu fields for the two conformal primaries in the massless multiplet \eqref{masslessmultipic2} which are going null.
Note that in all cases, the gauge parameter multiplets are always unitary un-shortened multiplets.

\section{AdS$_4$ Action and SUSY Transformations\label{AdSsection}}

In this section we construct the explicit action and SUSY transformation rules for the partially massless multiplet shown in \eqref{PMspin2multpice}. This is the simplest multiplet that includes a partially massless spin-2, as all others containing a partially massless spin-2 will also contain higher-spin particles.  For the sake of generality, and to study how the degrees of freedom behave as one approaches the partially massless limit, it is useful to begin by constructing the full AdS$_4$ massive gravity multiplet with spins $(2,3/2,3/2,1)$ and a generic graviton mass, described by the generic multiplet \eqref{genericlongpice} with $s=3/2$.\footnote{The flat-space version of this super-multiplet was studied in \cite{Buchbinder:2002gh}.} 

The free action is simply given by the sum of the free actions of a massive graviton $h_{\mu\nu}$, two massive Majorana gravitini $\Psi_{\mu}$ and $\Phi_{\mu}$, and a massive pseudo-vector\footnote{That $B_{\mu}$ must be a pseudo-vector for the fields to make a super-multiplet can be seen by analyzing the massless limit, in which the vector's longitudinal mode becomes the pseudo-scalar of a Wess--Zumino multiplet. We will make this explicit below.} $B_{\mu}$,
\beq\bal \label{eq:general action}
S&=\int d^4x\sqrt{-g}\bigg[-\frac{1}{2}\,\nabla^{\rho}h^{\mu\nu}\nabla_{\rho}h_{\mu\nu}+\nabla_{\rho}h^{\mu\nu}\nabla_{\mu}h^{\rho}_{\phantom{\rho}\nu}-\nabla_{\mu}h\nabla_{\nu}h^{\mu\nu}+\frac{1}{2}\,\nabla^{\mu}h\nabla_{\mu}h\\
&\quad-\frac{3}{L^2}\left(h^{\mu\nu}h_{\mu\nu}-{1\over 2}h^2\right) -\frac{m_2^2}{2}\left(h^{\mu\nu}h_{\mu\nu}-h^2\right)\\
&\quad-\frac{1}{2}\,\bar{\Psi}_{\mu}\gamma^{\mu\nu\rho}\nabla_{\nu}\Psi_{\rho}+\frac{\mu}{2}\,\bar{\Psi}_{\mu}\gamma^{\mu\nu}\Psi_{\nu} -\frac{1}{2}\,\bar{\Phi}_{\mu}\gamma^{\mu\nu\rho}\nabla_{\nu}\Phi_{\rho}+\frac{\mu'}{2}\,\bar{\Phi}_{\mu}\gamma^{\mu\nu}\Phi_{\nu}\\
&\quad-\frac{1}{4}\,G^{\mu\nu}G_{\mu\nu}-\frac{m_1^2}{2}\,B^{\mu}B_{\mu}\bigg]\,.
\eal\eeq
Here $g_{\mu\nu}$ is the metric of the AdS$_4$ background space of radius $L$, and covariant derivatives and gamma matrices are defined with respect to this background (see appendix \ref{app:ads conventions} for our conventions on spinors and covariant derivatives in AdS$_4$). We use the notations $h\equiv g^{\mu\nu}h_{\mu\nu}$ and $G_{\mu\nu}\equiv\nabla_{\mu}B_{\nu}-\nabla_{\nu}B_{\mu}$. 
There are four mass scales appearing in this action: $m_1$, $m_2$, $\mu$ and $\mu'$.  The scales $m_1$ and $m_2$ correspond to the vector and graviton masses.  The physical gravitino masses are related to the scales $\mu$ and $\mu'$ via
\beq \label{eq:gravitino masses}
m_{3/2}^2=\mu^2-\frac{1}{L^2}\,,\qquad m_{3/2}^{\prime\,2}=\mu^{\prime\,2}-\frac{1}{L^2}\,.
\eeq

We determine the SUSY transformation rules by writing down a fully generic ansatz for the symmetry transformation depending on a single infinitesimal Majorana fermion parameter $\epsilon$ and demanding invariance of the action. The SUSY parameter $\epsilon$ must be an AdS Killing spinor satisfying
\beq \label{eq:ads killing spinor eq}
\nabla_{\mu}\epsilon=\frac{1}{2L}\,\gamma_{\mu}\epsilon\,.
\eeq
This procedure fixes the vector and gravitino masses in terms of $m_2$ and the AdS radius as
\beq \label{eq:full masses}
m_1^2=m_2^2+\frac{2}{L^2}\,,\qquad \mu=\frac{1}{2L}-\sqrt{m_2^2+\frac{9}{4L^2}}\,,\qquad \mu'=\frac{1}{2L}+\sqrt{m_2^2+\frac{9}{4L^2}}\,.
\eeq
Note that $\mu'=-\mu$ in the flat space limit, yielding a Dirac-type mass term for the gravitini as required by the $R$-symmetry of Poincar\'e SUSY \cite{Zinoviev:2002xn}. Notice also that, given that the theory is symmetric upon interchanging the two gravitini and their masses, we are free to exchange the above solutions for $\mu$ and $\mu'$; we settle this ambiguity by choosing the field $\Psi_{\mu}$ to be the super-partner of the graviton in the limit $m_2\to0$ when $\Psi_{\mu}$ is also massless according to \eqref{eq:gravitino masses}.

On the other hand, the SUSY transformation is not uniquely determined solely from the invariance of the free action. This is not surprising as free theories are not subject to the Haag--{\L}opuzsa\'nski--Sohnius theorem \cite{Haag:1974qh}, and hence they may exhibit fermionic symmetries that are not supersymmetries. To find the true SUSY transformation law we must further impose the supersymmetry algebra, that is, we demand that two symmetry transformations should close to an AdS isometry,\footnote{As usual in on-shell formulations of SUSY, when applied to the spinors the algebra closes only upon use of the equations of motion.}
\beq \label{eq:ads susy algebra}
[\delta_{\epsilon_1},\delta_{\epsilon_2}]={\cal L}_{\xi}\,.
\eeq
Here 
${\cal L}_{\xi}$ denotes the Lie derivative with respect to the AdS Killing vector $\xi^{\mu}$, itself a function of the Majorana parameters $\epsilon_{1,2}$ of the symmetry transformations. This requirement fixes all the remaining coefficients in our general ansatz for the symmetry. The final result is
\beq\bal \label{eq:full symmetry}
\delta_{\epsilon} h_{\mu\nu}&=\Big[\bar{\Psi}_{(\mu}\gamma_{\nu)}+a_1\bar{\Phi}_{(\mu}\gamma_{\nu)}\Big]\epsilon+2L\nabla_{(\mu}\Big[a_2\bar{\Psi}_{\nu)}\epsilon+a_3\bar{\Phi}_{\nu)}\epsilon\Big]\,,\\
\delta_{\epsilon}\Psi_{\mu}&=\Big[\gamma^{\rho\sigma}\nabla_{\rho}h_{\sigma\mu}+b_1G_{\rho\sigma}\gamma^{\rho\sigma}\gamma_{\mu}i\gamma_5+b_2\gamma^{\nu}i\gamma_5(\nabla_{\mu}B_{\nu}+\nabla_{\nu}B_{\mu})\Big]\epsilon\\
&\quad+L\nabla_{\mu}\Big[b_3G_{\nu\rho}\gamma^{\nu\rho}i\gamma_5\epsilon\Big] +\frac{1}{L}\Big[b_4h_{\mu\nu}\gamma^{\nu}+b_5B_{\mu}i\gamma_5+b_6B^{\nu}\gamma_{\mu\nu}i\gamma_5\Big]\epsilon\,,\\
\delta_{\epsilon}\Phi_{\mu}&=\Big[a_1\gamma^{\rho\sigma}\nabla_{\rho}h_{\sigma\mu}+c_1G_{\rho\sigma}\gamma^{\rho\sigma}\gamma_{\mu}i\gamma_5+c_2\gamma^{\nu}i\gamma_5(\nabla_{\mu}B_{\nu}+\nabla_{\nu}B_{\mu})\Big]\epsilon\\
&\quad+L\nabla_{\mu}\Big[c_3G_{\nu\rho}\gamma^{\nu\rho}i\gamma_5\epsilon\Big] +\frac{1}{L}\Big[c_4h_{\mu\nu}\gamma^{\nu}+c_5B_{\mu}i\gamma_5+c_6B^{\nu}\gamma_{\mu\nu}i\gamma_5\Big]\epsilon\,,\\
\delta_{\epsilon} B_{\mu}&=\Big[d_1\bar{\Psi}_{\mu}i\gamma_5+d_2\bar{\Phi}_{\mu}i\gamma_5\Big]\epsilon\,.\\
\eal\eeq
The coefficients $a$, $b$, $c$ and $d$ that appear in these expressions are given explicitly by
\beq\nonumber
a_1=\frac{\sqrt{9+4x^2}-3}{2x}\,,\qquad a_2=\frac{\sqrt{9+4x^2}-3}{4x^2}\,,\qquad a_3=-\frac{1}{2x}\,,
\eeq
\beq\nonumber
b_1=-\frac{\sqrt{9+4x^2}-4}{4\sqrt{6}\,x}\,,\qquad b_2=-\frac{\sqrt{9+4x^2}-1}{2\sqrt{6}\,x}\,,\qquad b_3=-\frac{1}{2\sqrt{6}\,x}\,,
\eeq
\beq\nonumber
b_4=\frac{\sqrt{9+4x^2}-1}{2}\,,\qquad b_5=\frac{3\sqrt{9+4x^2}-(7+2x^2)}{\sqrt{6}\,x}\,,\qquad b_6=-\frac{\sqrt{9+4x^2}-(3+x^2)}{\sqrt{6}\,x}\,,
\eeq
\beq\nonumber
c_1=\frac{\sqrt{9+4x^2}-(3-4x^2)}{8\sqrt{6}\,x^2}\,,\qquad c_2=-\frac{\sqrt{9+4x^2}-(3+2x^2)}{2\sqrt{6}\,x^2}\,,\qquad c_3=-\frac{\sqrt{9+4x^2}-3}{4\sqrt{6}\,x^2}\,,
\eeq
\beq\nonumber
c_4=\frac{\sqrt{9+4x^2}-(3+2x^2)}{2x}\,,\quad c_5=\frac{(1-x^2)\sqrt{9+4x^2}-3(1+x^2)}{\sqrt{6}\,x^2}\,,\quad c_6=\frac{\sqrt{9+4x^2}+1}{2\sqrt{6}}\,,
\eeq
\beq\nonumber
d_1=-\sqrt{\frac{3}{2}}\,\frac{\sqrt{9+4x^2}-3}{2x}\,,\qquad d_2=\sqrt{\frac{3}{2}}\,,
\eeq
where $x\equiv m_2L\,$.

With the normalization chosen in \eqref{eq:full symmetry} for the symmetry parameter $\epsilon$, the SUSY algebra is realized as in \eqref{eq:ads susy algebra} with
\beq
\xi^{\mu}(\epsilon_1,\epsilon_2)=\frac{\sqrt{9+4L^2m_2^2}\left(3-\sqrt{9+4L^2m_2^2}\right)}{L^2m_2^2}\,(\bar{\epsilon}_1\gamma^{\mu}\epsilon_2)\,.
\eeq
That $\xi^{\mu}$ is an AdS Killing vector follows from the fact that $\epsilon_{1,2}$ are Killing spinors satisfying \eqref{eq:ads killing spinor eq}.

\subsection{Flat limit}

Before considering the partially massless case it is instructive to first look at the special limits $L\to\infty$ and $m_2\to0$ which, in addition to being physically interesting, are simpler to understand. The flat limit is straightforward as nothing special occurs to the degrees of freedom and no divergences happen at the level of the SUSY transformation. When $L\to\infty$ Eq.\ \eqref{eq:full symmetry} becomes
\beq\bal \label{eq:susy flat limit}
\delta_{\epsilon} h_{\mu\nu}&=\Big[\bar{\Psi}_{(\mu}\gamma_{\nu)}+\bar{\Phi}_{(\mu}\gamma_{\nu)}\Big]\epsilon+\frac{1}{m_2}\,\partial_{(\mu}\Big[\bar{\Psi}_{\nu)}-\bar{\Phi}_{\nu)}\Big]\epsilon\,,\\
\delta_{\epsilon}\Psi_{\mu}&=\Big[\gamma^{\rho\sigma}\partial_{\rho}h_{\sigma\mu}-\frac{1}{2\sqrt{6}}\,G_{\rho\sigma}\gamma^{\rho\sigma}\gamma_{\mu}i\gamma_5-\frac{1}{\sqrt{6}}\,\gamma^{\nu}i\gamma_5(\partial_{\mu}B_{\nu}+\partial_{\nu}B_{\mu})\Big]\epsilon\\
&\quad-\frac{1}{2\sqrt{6}\,m_2}\,\partial_{\mu}G_{\nu\rho}\gamma^{\nu\rho}i\gamma_5\epsilon +m_2\Big[h_{\mu\nu}\gamma^{\nu}-\frac{2}{\sqrt{6}}\,B_{\mu}i\gamma_5+\frac{1}{\sqrt{6}}\,B^{\nu}\gamma_{\mu\nu}i\gamma_5\Big]\epsilon\,,\\
\delta_{\epsilon}\Phi_{\mu}&=\Big[\gamma^{\rho\sigma}\partial_{\rho}h_{\sigma\mu}+\frac{1}{2\sqrt{6}}\,G_{\rho\sigma}\gamma^{\rho\sigma}\gamma_{\mu}i\gamma_5+\frac{1}{\sqrt{6}}\,\gamma^{\nu}i\gamma_5(\partial_{\mu}B_{\nu}+\partial_{\nu}B_{\mu})\Big]\epsilon\\
&\quad-\frac{1}{2\sqrt{6}\,m_2}\,\partial_{\mu}G_{\nu\rho}\gamma^{\nu\rho}i\gamma_5\epsilon -m_2\Big[h_{\mu\nu}\gamma^{\nu}+\frac{2}{\sqrt{6}}\,B_{\mu}i\gamma_5-\frac{1}{\sqrt{6}}\,B^{\nu}\gamma_{\mu\nu}i\gamma_5\Big]\epsilon\,,\\
\delta_{\epsilon} B_{\mu}&=-\sqrt{\frac{3}{2}}\,\Big[\bar{\Psi}_{\mu}i\gamma_5-\bar{\Phi}_{\mu}i\gamma_5\Big]\epsilon\,.\\
\eal\eeq
This is the unitary gauge version of the supersymmetry of the flat-space massive gravity multiplet constructed in \cite{Zinoviev:2002xn} using the St\"uckelberg formulation. From \eqref{eq:gravitino masses} and \eqref{eq:full masses} we see that the masses all become equal in this limit, as expected, while the Killing spinor condition \eqref{eq:ads killing spinor eq} now simply implies that the SUSY parameter $\epsilon$ is a constant Majorana spinor.

\subsection{Massless limit\label{masslessadslimitsection}}

More care is needed to treat the massless graviton limit since the symmetry rule \eqref{eq:full symmetry} is singular at $m_2=0$. This is a manifestation of the fact that the number of degrees of freedom is discontinuous in this limit as both $h_{\mu\nu}$ and $\Psi_{\mu}$ become massless. We can handle this by means of a St\"uckelberg replacement, introducing a vector $A_{\mu}$ that restores the linearized diff invariance of a massless spin-2 on AdS$_4$, and a Majorana spinor $\lambda$ that restores the gauge symmetry of a massless spin-3/2 on AdS$_4$,
\beq\bal \label{eq:massless stuck}
h_{\mu\nu}&\longrightarrow h_{\mu\nu}+\frac{1}{\sqrt{2}\,m_2}\left(\nabla_{\mu}A_{\nu}+\nabla_{\nu}A_{\mu}\right)\,,\\
\Psi_{\mu}&\longrightarrow \Psi_{\mu}+\frac{1}{m_2}\left(\nabla_{\mu}\lambda-\frac{1}{2L}\,\gamma_{\mu}\lambda\right)\,.
\eal\eeq
The prefactors are chosen so that the fields $A_{\mu}$ and $\lambda$ become canonically normalized upon taking the limit $m_2\to0$. The final action reads
\beq\bal
S_{(m_2=0)}&=\int d^4x\sqrt{-g}\bigg[-\frac{1}{2}\,\nabla^{\rho}h^{\mu\nu}\nabla_{\rho}h_{\mu\nu}+\nabla_{\rho}h^{\mu\nu}\nabla_{\mu}h^{\rho}_{\phantom{\rho}\nu}-\nabla_{\mu}h\nabla_{\nu}h^{\mu\nu}+\frac{1}{2}\,\nabla^{\mu}h\nabla_{\mu}h\\
&\quad-\frac{3}{L^2}\left(h^{\mu\nu}h_{\mu\nu}-\frac{h^2}{2}\right) \\
&\quad -\frac{1}{2}\,\bar{\Psi}_{\mu}\gamma^{\mu\nu\rho}\nabla_{\nu}\Psi_{\rho}-\frac{1}{2L}\,\bar{\Psi}_{\mu}\gamma^{\mu\nu}\Psi_{\nu}-\frac{1}{2}\,\bar{\Phi}_{\mu}\gamma^{\mu\nu\rho}\nabla_{\nu}\Phi_{\rho}+\frac{1}{L}\,\bar{\Phi}_{\mu}\gamma^{\mu\nu}\Phi_{\nu}\\
&\quad-\frac{1}{4}\,F^{\mu\nu}F_{\mu\nu}-\frac{3}{L^2}\,A^{\mu}A_{\mu}-\frac{1}{4}\,G^{\mu\nu}G_{\mu\nu}-\frac{1}{L^2}\,B^{\mu}B_{\mu}-\frac{1}{2}\,\bar{\lambda}\gamma^{\mu}\nabla_{\mu}\lambda+\frac{1}{L}\,\bar{\lambda}\lambda \bigg]\,,
\eal\eeq
where $F_{\mu\nu}\equiv \nabla_\mu A_\nu-\nabla_\nu A_\mu.$  This action has the expected gauge symmetries,
\beq
\delta_{\xi}h_{\mu\nu}=\nabla_{\mu}\xi_{\nu}+\nabla_{\nu}\xi_{\mu}\,,\qquad \delta_{\eta}\Psi_{\mu}=\nabla_{\mu}\eta-\frac{1}{2L}\,\gamma_{\mu}\eta\,,
\eeq
of massless spin-2 and massless spin-3/2 fields in AdS$_4$, which now form an independent massless short super-multiplet $(2,3/2)$, as depicted in figure \eqref{masslessmultipic2} with $s=3/2$.

The rest of the fields have masses given by
\beq
m_\Phi^2=\frac{3}{L^2}\,,\qquad m_A^2=\frac{6}{L^2}\,,\qquad m_B^2=\frac{2}{L^2}\,,\qquad m_\lambda^2=\frac{4}{L^2}\,,
\eeq
and make their own massive AdS gravitino multiplet $\left\{ 1\right\}_{3}$, with the massive spins $(3/2,1,1,1/2)$.   One can check this by comparing with the mass spectrum of a generic gravitino multiplet, which we describe in Appendix \ref{adsgravitinoappendix}.  We can confirm this by performing the St\"uckelberg replacement \eqref{eq:massless stuck} in the full SUSY transformation \eqref{eq:full symmetry}, and then take the limit $m_2\to0$. One then finds that the two sets of fields, $(h_{\mu\nu},\Psi_{\mu})$ and $(\Phi_{\mu},A_\mu,B_\mu,\lambda)$, indeed transform independently of each other under supersymmetry:
\beq\bal \label{eq:susy massless limit graviton}
\delta_{\epsilon} h_{\mu\nu}&=\bar{\Psi}_{(\mu}\gamma_{\nu)}\epsilon\,,\\
\delta_{\epsilon}\Psi_{\mu}&=\gamma^{\rho\sigma}\nabla_{\rho}h_{\sigma\mu}\epsilon +\frac{1}{L}\,h_{\mu\nu}\gamma^{\nu}\epsilon\,,\\
\eal\eeq
\beq\bal \label{eq:susy massless limit gravitino}
\delta_{\epsilon'}\Phi_{\mu}&=\Big[-\frac{1}{12}\,F_{\rho\sigma}\gamma^{\rho\sigma}\gamma_{\mu}-\frac{2}{3}\,\gamma^{\nu}(\nabla_{\mu}A_{\nu}+\nabla_{\nu}A_{\mu})\Big]\epsilon'+\frac{L}{6}\,\nabla_{\mu}\Big[F_{\nu\rho}\gamma^{\nu\rho}\epsilon'\Big]-\frac{2}{3L}\,A^{\nu}\gamma_{\mu\nu}\epsilon'\,,\\
&\quad+\Big[\frac{7}{12\sqrt{3}}\,G_{\rho\sigma}\gamma^{\rho\sigma}\gamma_{\mu}i\gamma_5+\frac{2}{3\sqrt{3}}\,\gamma^{\nu}i\gamma_5(\nabla_{\mu}B_{\nu}+\nabla_{\nu}B_{\mu})\Big]\epsilon'\\
&\quad-\frac{L}{6\sqrt{3}}\,\nabla_{\mu}\Big[G_{\nu\rho}\gamma^{\nu\rho}i\gamma_5\epsilon'\Big]+\frac{1}{L}\Big[-\frac{16}{3\sqrt{3}}\,B_{\mu}i\gamma_5+\frac{2}{\sqrt{3}}\,B^{\nu}\gamma_{\mu\nu}i\gamma_5\Big]\epsilon'\,,\\
\delta_{\epsilon'} A_{\mu}&=-\bar{\Phi}_{\mu}\epsilon'+\bar{\lambda}\gamma_{\mu}\epsilon'+\frac{L}{3}\,\nabla_{\mu}\left[\bar{\lambda}\epsilon'\right]\,,\\
\delta_{\epsilon'} B_{\mu}&=\sqrt{3}\,\bar{\Phi}_{\mu}i\gamma_5\epsilon'-\frac{1}{\sqrt{3}}\,\bar{\lambda}\gamma_{\mu}i\gamma_5\epsilon'-\frac{L}{\sqrt{3}}\,\nabla_{\mu}\left[\bar{\lambda}i\gamma_5\epsilon'\right]\,,\\
\delta_{\epsilon'}\lambda&=\frac{1}{2}\Big[F_{\mu\nu}\gamma^{\mu\nu}+\frac{4}{L}\,A^{\mu}\gamma_{\mu}\Big]\epsilon'-\frac{1}{2\sqrt{3}}\Big[G_{\mu\nu}\gamma^{\mu\nu}i\gamma_5+\frac{4}{L}\,B^{\mu}\gamma_{\mu}i\gamma_5\Big]\epsilon'\,.
\eal\eeq
We have written the Majorana parameter as $\epsilon'$ in the transformation of the gravitino multiplet to emphasize that it is now independent from that of the massless graviton multiplet. Comparing Eqs.\ \eqref{eq:susy massless limit gravitino} and \eqref{eq:gravitino multiplet susy} one can verify that they indeed match for the value $m_{3/2}^2={3}/{L^2}$ of the gravitino mass quoted above, thus confirming that it is an AdS SUSY multiplet.

In terms of the SUSY representations, what is described here is the branching rule $\left\{ 3\over 2 \right\}_{\Delta}\underset{\Delta \rightarrow {5\over 2}}{\rightarrow}\left\{ 3\over 2 \right\}_{5\over 2} \oplus \left\{ 1\right\}_{3}$, as described in Section \ref{branchingsection}.

\subsection{Partially massless limit\label{PMadssection}}

We now examine the massive gravity multiplet when the spin-2 is partially massless, by taking the mass $m_2^2=-2/L^2$. From \eqref{eq:full masses} we get the other masses,
\beq
m_{3/2}^2=-\frac{1}{L^2}\,,\qquad m_{3/2}^{\prime2}=0\,,\qquad m_1^2=0\,,
\eeq
and hence the vector and one of the gravitini are massless at the PM point. This implies that if we take the PM mass value as a limit starting with the generic theory then the number degrees of freedom will be discontinuous, just like in the massless case considered above. However, what is different here is that the number of bosonic and fermionic degrees of freedom would still match if one were to ignore the St\"uckelberg fields. This means that it is consistent, from the point of view of the supersymmetry, to directly set the PM value $m_2^2=-2/L^2$ and ignore the lost degrees of freedom (which are therefore expected to form their own independent SUSY multiplet, as we will check below).

This is also confirmed by noting that the coefficients of the full SUSY transformation law, eq.\ \eqref{eq:full symmetry}, are all finite at the PM point. However, because of the appearance of $m_2=i\sqrt{2}/L$, some coefficients turn out to be imaginary. This manifests the fact that some of the fields in the multiplet carry negative norm, whereas in writing the original action as in \eqref{eq:general action} we implicitly assumed that all norms were positive. The relative signs of the norms can be inferred by looking back at eqs.\ \eqref{eq:norm1}, \eqref{eq:norm2}, \eqref{null2levelde} and \eqref{eq:norm4}, where we take $s=3/2$ and $\Delta=3/2$ for the PM multiplet. The field that corresponds to the conformal primary is $\Psi_{\mu}$, which we take to have positive norm at the level of the action, whose overall sign is irrelevant. It follows that $B_{\mu}$ and $\Phi_{\mu}$ must have negative norm, while $h_{\mu\nu}$ must have positive norm. We therefore perform the replacement $B_{\mu}\to iB_{\mu}$ and $\Phi_{\mu}\to i\Phi_{\mu}$ both in the action and in the SUSY transformation. For the final PM spin-2 multiplet action we get
\beq\bal \label{eq:pm action}
S_{\mathrm{PM}}&=\int d^4x\sqrt{-g}\bigg[-\frac{1}{2}\,\nabla^{\rho}h^{\mu\nu}\nabla_{\rho}h_{\mu\nu}+\nabla_{\rho}h^{\mu\nu}\nabla_{\mu}h^{\rho}_{\phantom{\rho}\nu}-\nabla_{\mu}h\nabla_{\nu}h^{\mu\nu}+\frac{1}{2}\,\nabla^{\mu}h\nabla_{\mu}h\\
&\quad -\frac{2}{L^2}\left(h^{\mu\nu}h_{\mu\nu}-\frac{1}{4}\,h^2\right) \\
&\quad+\frac{1}{4}\,G^{\mu\nu}G_{\mu\nu}-\frac{1}{2}\,\bar{\Psi}_{\mu}\gamma^{\mu\nu\rho}\nabla_{\nu}\Psi_{\rho}+\frac{1}{2}\,\bar{\Phi}_{\mu}\gamma^{\mu\nu\rho}\nabla_{\nu}\Phi_{\rho}-\frac{1}{2L}\,\bar{\Phi}_{\mu}\gamma^{\mu\nu}\Phi_{\nu}\bigg]\,.
\eal\eeq
The theory is invariant under the gauge symmetries
\beq\bal \label{eq:pm gauge sym}
\delta_{\xi} h_{\mu\nu}&=\nabla_{\mu}\nabla_{\nu}\xi-\frac{1}{L^2}\,g_{\mu\nu}\xi\,,\\
\delta_{\eta} \Phi_{\mu}&=\nabla_{\mu}\eta+\frac{1}{2L}\,\gamma_{\mu}\eta\,,\\
\delta_{\alpha} B_{\mu}&=\nabla_{\mu}\alpha\,,
\eal\eeq
and under the supersymmetry
\beq\bal \label{eq:pm supersymmetry}
\delta_{\epsilon}{h}_{\mu\nu}&=\bigg[\bar{\Psi}_{(\mu}\gamma_{\nu)}+\frac{1}{\sqrt{2}}\,\bar{\Phi}_{(\mu}\gamma_{\nu)}\bigg]\epsilon+L\nabla_{(\mu}\bigg[\frac{1}{2}\,\bar{\Psi}_{\nu)}\epsilon+\frac{1}{\sqrt{2}}\,\bar{\Phi}_{\nu)}\epsilon\bigg]\,,\\
\delta_{\epsilon}\Psi_{\mu}&=\bigg[\gamma^{\rho\sigma}\nabla_{\rho}{h}_{\sigma\mu}-\frac{\sqrt{3}}{8}\,G_{\rho\sigma}\gamma^{\rho\sigma}\gamma_{\mu}i\gamma_5\bigg]\epsilon+L\nabla_{\mu}\bigg[\frac{1}{4\sqrt{3}}\,G_{\nu\rho}\gamma^{\nu\rho}i\gamma_5\epsilon\bigg]\,, \\
\delta_{\epsilon}\Phi_{\mu}&=\bigg[-\frac{1}{\sqrt{2}}\,\gamma^{\rho\sigma}\nabla_{\rho}{h}_{\sigma\mu}+\frac{5}{8\sqrt{6}}\,G_{\rho\sigma}\gamma^{\rho\sigma}\gamma_{\mu}i\gamma_5+\frac{1}{2\sqrt{6}}\,\gamma^{\nu}i\gamma_5(\nabla_{\mu}B_{\nu}+\nabla_{\nu}B_{\mu})\bigg]\epsilon\\
&\quad+L\nabla_{\mu}\bigg[-\frac{1}{4\sqrt{6}}\,G_{\nu\rho}\gamma^{\nu\rho}i\gamma_5\epsilon\bigg] +\frac{1}{L}\bigg[\frac{1}{\sqrt{2}}\,{h}_{\mu\nu}\gamma^{\nu}-\frac{\sqrt{3}}{\sqrt{2}}\,B_{\mu}i\gamma_5+\frac{1}{\sqrt{6}}\,B^{\nu}\gamma_{\mu\nu}i\gamma_5\bigg]\epsilon\,,\\
\delta_{\epsilon}B_{\mu}&=\bigg[\frac{\sqrt{3}}{2}\,\bar{\Psi}_{\mu}i\gamma_5+\frac{\sqrt{3}}{\sqrt{2}}\,\bar{\Phi}_{\mu}i\gamma_5\bigg]\epsilon\,.\\
\eal\eeq
This SUSY transformation is now consistent with the reality properties of the fields (which can be seen explicitly for instance by using a ``really real'' representation where the gamma matrices and Majorana fields are all purely real, so that $\gamma_5$ is purely imaginary \cite{Freedman:2012zz}).

We have thus succeeded in performing an explicit construction of the simplest super-multiplet containing a partially massless spin-2. The result is fully consistent with the group theory analysis of the previous section. From the action \eqref{eq:pm action} we see that the relative signs of the norms agree with those of the conformal primary state and its descendants, while the final result for the SUSY transformation \eqref{eq:pm supersymmetry} confirms that no shortening occurs at the PM point. Finally, it is straightforward to also check that the mass spectrum agrees with the AdS/CFT relation \eqref{AdSCFTformpe} with the scaling dimensions as given in \eqref{PMspin2multpice}.

We remarked above that taking the PM limit starting from the generic massive multiplet does not require the introduction of St\"uckelberg fields as long as one is only interested in the PM multiplet and its symmetries. It is however instructive to anyway perform a St\"uckelberg replacement as a further consistency check, which will also serve us to justify the claim that the gauge parameters of the theory themselves form an AdS SUSY multiplet. Following the pattern in eq.\ \eqref{eq:pm gauge sym} we do the replacements
\beq\bal
{h}_{\mu\nu}&\longrightarrow {h}_{\mu\nu}+\frac{1}{\sqrt{3}}\,\frac{L^2}{\sqrt{2+L^2m_2^2}}\left(\nabla_{\mu}\nabla_{\nu}\varphi-\frac{1}{L^2}\,g_{\mu\nu}\varphi\right)\,,\\
\Phi_{\mu}&\longrightarrow\Phi_{\mu}+\frac{1}{\sqrt{3}}\,\frac{L}{\sqrt{2+L^2m_2^2}}\left(\nabla_{\mu}\lambda+\frac{1}{2L}\,\gamma_{\mu}\lambda\right)\,,\\
B_{\mu}&\longrightarrow B_{\mu}+\frac{L}{\sqrt{2+L^2m_2^2}}\,\nabla_{\mu}\pi\,.
\eal\eeq
Again the prefactors are such that the St\"uckelbergs $\varphi$, $\lambda$ and $\pi$ come out canonically normalized after one takes the limit $m_2^2\to-2/L^2$ from above. All three fields $\varphi$, $\pi$ and $\lambda$ have negative norm in this limit, as one can straightforwardly check by performing the replacement in the action. (Alternatively, one may approach the PM point from below and then all the St\"uckelbergs will have positive norm.) From the result one can read off the masses as
\beq \label{eq:pm stuck masses}
m_{\lambda}^2=\frac{4}{L^2}\,,\qquad m_{\varphi}^2=\frac{4}{L^2}\,,\qquad m_{\pi}^2=0\,.
\eeq
Replacing the St\"uckelberg fields in the SUSY transformation one finds, in the limit $m_2^2\to-2/L^2$, that they can indeed be isolated and satisfy an independent transformation rule,
\beq\bal \label{eq:pm stuck susy}
\delta_{\epsilon}\varphi&=\bar{\lambda}\epsilon\,,\qquad \delta_{\epsilon}\pi=\bar{\lambda}i\gamma_5\epsilon\,,\\
\delta_{\epsilon}\lambda&=\gamma^{\mu}(\nabla_{\mu}\varphi+i\gamma_5\nabla_{\mu}\pi)\epsilon-\frac{1}{L}\,\varphi\epsilon-\frac{3}{L}\,\pi i\gamma_5\epsilon\,.\\
\eal\eeq
In appendix \ref{app:other multiplets} we give the generic AdS Wess--Zumino multiplet. Comparison of \eqref{eq:pm stuck masses} and \eqref{eq:pm stuck susy} with \eqref{eq:wz multiplet masses} and \eqref{eq:wz multiplet susy} then verifies that the St\"uckelbergs $\varphi$, $\lambda$ and $\pi$ form a super-multiplet.

In terms of the SUSY representations, what is described here is the branching rule $\left\{ 3\over 2 \right\}_{\Delta}\underset{\Delta \rightarrow {3\over 2}}{\rightarrow}\left\{ 3\over 2 \right\}_{3\over 2} \oplus \left\{ 0\right\}_{3}$, as described in Section \ref{branchingsection}. In the action, the \stu fields carry these null states, and hence they split off into their own massive multiplet.  Note that, unlike in the massless limit, none of the conformal primaries themselves are becoming null, only descendants are.  This is why there was no obstruction to directly taking the limit in the action.

\section{Conclusions}

We have studied non-unitary representations of the ${\cal N}=1$ 3d superconformal algebra.  These correspond to non-unitary ${\cal N}=1$ supersymmetric multiplets on AdS$_4$.  Among these, we have identified the representations containing partially massless fields.  In addition, we have found supersymmetric BPS-like shortening conditions that occur among the non-unitary representations.  These shortening conditions have no analogue among the unitary representations, and include supersymmetric examples of exotic ``extended modules.''

We have identified the simplest multiplet which contains the partially massless spin 2.  This multiplet contains a massless vector, a massless spin 3/2 and a massive spin 3/2 in addition to the partially massless spin 2.  We have written the free AdS$_4$ action and supersymmetry transformations for this multiplet, and studied how it arises from the partially massless limit of the full massive supermultiplet.

There is still a possibility that an interacting theory of this partially massless SUSY multiplet exists.  This would be a partially massless supergravity, and it would have a consistent truncation to a bosonic part consisting of an interacting partially massless spin-2 and a photon.  This possibility is not yet ruled out by any of the no-go theorems against interacting partially massless spin-2 fields which are known so far, so it would be interesting to further study such a possibility.

We restricted our study of partially massless fields to those with standard quantization conditions, which meant that the new exotic non-unitary shortening conditions \eqref{short2multe}, \eqref{short3multe} and the extended multiplet at $\Delta=1$ did not play a role.   It would be interesting to understand how these other multiplets play out in AdS$_4$.
Finally, it would be interesting to extend this study to ${\cal N}=2$ and higher and classify the resulting partially massless multiplets of extended SUSY, as well as to go to higher dimensions where there are more possibilities including mixed symmetry partially massless representations.

\bigskip
{\bf Acknowledgements:}
The authors are grateful to Chris Brust, Clay Cordova, Frederik Denef, Austin Joyce, and David Poland for discussions and comments.  SGS is supported by the European Research Council under the European Community's Seventh Framework Programme (FP7/2007-2013 Grant Agreement no.\ 307934, NIRG project).  KH acknowledges support from DOE grant DE-SC0019143.  RAR is supported by DOE grant DE-SC0011941, NASA grant NNX16AB27G and Simons Foundation Award Number 555117.

\appendix

\section{$d=3$ spinor conventions\label{spinorappendix}}

Here we detail our conventions for $d=3$ euclidean spinors used in the superconformal algebra and its representations.   The space indices are $i,j,\ldots$ which range over $1,2,3$.  The metric is 
\be \delta_{ij}={\rm diag}(1,1,1),\ee
and is used to raise and lower space indices.  The fundamental spinor representation in $d=3$ euclidean space is two dimensional and is irreducible and complex (there is no Majorana or Weyl condition).  
We use $a,b,c,\ldots$ ranging over $1,2$ for spinor indices.  

The gamma matrices are the standard Pauli matrices $\sigma^i$,
\be \sigma^1=\left(\begin{array}{cc} & 1 \\1 & \end{array}\right), \ \ \sigma^2=\left(\begin{array}{cc} & -i \\i & \end{array}\right), \ \ \sigma^3=\left(\begin{array}{cc}1 &  \\ & -1\end{array}\right).
\ \ \ \ 
\ee
As determined by their transformation laws, they come with the index structure
\be \left(\sigma^i\right)^{a}_{\ b}\, . \label{gammaindexdefe}
\ee
They span the space $su(2)$ of traceless hermitian matrices.  They satisfy the standard gamma matrix anti-commutation rule,
\be \{\sigma_i,\sigma_j\}=2\delta_{ij}\, .\ee

The Lorentz generators are given by 
\be M_{ij}={1\over 2}\sigma_{ij},\ \ \ \sigma^{ij}\equiv \sigma^{[i}\sigma^{j]}\, ,\ee
which satisfy
\be \left[M_{ij},\sigma_k\right]=-\delta_{ik}\sigma_j+\delta_{jk}\sigma_i .\ee
We have the following relation,
\be \sigma^i\sigma^j=\delta^{ij}+i\epsilon^{ij}_{\ \  k}\sigma^k,\ee
where $\epsilon_{ijk}$ is the $d=3$ anti-symmetric symbol (with the convention $\epsilon_{123}=1$).
This implies
\be \sigma^{ij}=i\epsilon^{ij}_{\ \  k}\sigma^k\, ,\ee
so the Lorentz generators are the same as the gamma matrices (this is the Lie algebra isomorphism $su(2)\simeq so(3)$).

Useful traces of the gammas are
\bea  
&& \Tr\left[\sigma^i\sigma^j\right]=2\delta^{ij}\, , \\
&& \Tr\left[\sigma^i\sigma^j\sigma^k \right]=2i\epsilon^{ijk}\, ,
\eea
and they satisfy a completeness relation
\be \left(\sigma^i\right)^a_{\ b}\left(\sigma_i\right)^c_{\ d}=2\delta^a_{\ d}\delta^{c}_{\ b}-\delta^a_{\ b}\delta^{c}_{\ d}\, ,\ee
 \be \left(\sigma^{ij}\right)^a_{\ b}\left(\sigma_{ij}\right)^c_{\ d}=-4\delta^a_{\ d}\delta^{c}_{\ b}+2\delta^a_{\ b}\delta^{c}_{\ d}.\ee

Indices on spinors are raised and lowered with the invariant tensor $\epsilon$,
\be \epsilon_{ab}= \epsilon^{ab}=\left(\begin{array}{cc} & 1 \\-1 &  \end{array}\right),\ \ \ \ \epsilon^{ac}\epsilon_{bc}=\delta_{\ b}^a.\ee
For a spinor $\psi$ we raise and lower spinor indices using the convention
\be \psi_a=\epsilon _{ab}\psi^b\, ,\ \ \ \ \psi^a= \psi_b\epsilon^{ba}.\ee
Note that we defined the $\epsilon$'s so that we have the property $\epsilon^{ab}=\epsilon_{cd}\epsilon^{ca}\epsilon^{db}$ i.e. $\epsilon^{ab}$ is in fact the same as $\epsilon_{ab}$ with raised indices.  We only have to be careful that $\delta$ has indices in the right position; the Kronecker delta is $\delta^a_{\ b}=\epsilon^a_{\ b}$, and the raised and lowered version is $\delta^{\ a}_{b}=-\delta^a_{\ b}$.  With these rules the raising and lowering of indices on $\epsilon$'s and $\delta$'s is consistent.  For any tensor $T$ we have 
\be T^{a\ \ \ \cdots}_{\ \ a\ \ \cdots}=-T_{a\ \ \  \cdots}^{\ a\  \ \ \cdots }\ \ \ .\label{raiseloweape}\ee

There are identities that follow from the fact that we are on a two dimensional space: any pair of anti-symmetric indices can be replaced by an $\epsilon$,
\be T^{[ab]\ \cdots}_{\ \ \ \ \ \ldots}={1\over 2}\epsilon^{ab}T^{c\ \ \ \cdots}_{\ \ c\  \ \ldots},\label{antisymsqeide}\ee
and the anti-symmetrization of any three indices vanishes,
\be T^{[abc]\ \cdots}_{\ \ \ \ \ \ \ldots}=0\, .\ee

Raising an index on the gamma matrices \eqref{gammaindexdefe}, the result is symmetric\footnote{Explicitly:
\be \sigma^1\epsilon=\left(\begin{array}{cc}-1 & 0 \\ 0 & 1\end{array}\right),\ \ \ \ \sigma^2\epsilon=\left(\begin{array}{cc}i & 0 \\0 & i\end{array}\right),\ \ \ \ \sigma^3\epsilon=\left(\begin{array}{cc}0 & 1 \\1 & 0\end{array}\right).\ \ \ \ \label{gammaupexe}
\ee
}:
\be \left(\sigma^i\right)^{ab}=\left(\sigma^i\right)^{a}_{\ c}\epsilon^{cb},\ \ \ \ \left(\sigma^i\right)^{ab}=\left(\sigma^i\right)^{ba}.\label{raisedgammadefse}\ee

We have the complex conjugation properties
\be \left.\left(\sigma_i\right)^a_{\ b}\right.^\ast =\left(\sigma_i\right)_{a}^{\ b},\ \ \  \left.\left(\sigma_i\right)^{ab}_{\ }\right.^\ast =-\left(\sigma_i\right)_{ab}^{\ },\ \ \  \left.\left(\sigma_i\right)^{}_{ab}\right.^\ast =-\left(\sigma_i\right)_{}^{ab },\ \ \ \ \left.\left(\sigma_i\right)_a^{\ b}\right.^\ast =\left(\sigma_i\right)^{a}_{\ b}, \ee
\be \left.\left(\sigma_{ij}\right)^a_{\ b}\right.^\ast =-\left(\sigma_{ij}\right)_{a}^{\ b},\ \ \  \left.\left(\sigma_{ij}\right)^{ab}_{\ }\right.^\ast =\left(\sigma_{ij}\right)_{ab}^{\ }, \ \ \  \left.\left(\sigma_{ij}\right)^{}_{ab}\right.^\ast =\left(\sigma_{ij}\right)_{}^{ab },\ \ \ \ \left.\left(\sigma_{ij}\right)_a^{\ b}\right.^\ast =-\left(\sigma_{ij}\right)^{a}_{\ b}. \ee

\subsection{Representations}

The irreducible representations of $SU(2)$ can be packaged as symmetric spinors,
\be T^{a_1a_2\cdots a_{2s}},\ee
where $s=0,1/2,1,3/2,\ldots$ is a half integer which corresponds to the spin. These are the only irreducible representations; any lowered indices are equivalent to upper ones by raising with $\epsilon^{ab}$, and any non-symmetric components can be reduced to fully symmetric ones by using \eqref{antisymsqeide} to remove non-symmetric parts.  Note that there is no tracelessness condition imposed on these tensors, since there is no invariant trace operation.

The irreducible representations of $SO(3)$, on the other hand, are symmetric {\it traceless} tensors
\be T^{i_1\cdots i_s}\,,\ee
with $s=0,1,2,\ldots$ an integer.  All other symmetry types can be reduced to this by using the epsilon symbol $\epsilon_{ijk}$, and traces removed using $\delta_{ij}$.

The $SO(3)$ tensor representations of rank $s$ are equivalent to the integer spin $SU(2)$ spinor representations of rank $2s$. 
We can use the gamma matrices \eqref{raisedgammadefse} to pass between them,
\be T^{a_1b_1\cdots a_s b_s}=\sigma_{i_1}^{a_1b_1}\cdots \sigma_{i_s}^{a_1b_s}T^{i_1\cdots i_s}\, , \ee
\be T^{i_1\cdots i_s}={1\over 2^s}\sigma^{i_1}_{a_1b_1}\cdots \sigma^{i_s}_{a_1b_s} T^{a_1b_1\cdots a_s b_s}\, .\ee 
The odd rank spinor representations have no tensor counterpart.

\section{$D=4$ AdS spinor conventions} \label{app:ads conventions}

We follow the conventions of \cite{Freedman:2012zz} for spinors in $D=4$ AdS space. The covariant gamma matrices are given by $\gamma_{\mu}\equiv e^{\hat{\mu}}_{\phantom{\hat{\mu}}\mu}\gamma_{\hat{\mu}}$, where the $\gamma_{\hat{\mu}}$ are the standard constant gamma matrices and $e^{\hat{\mu}}_{\phantom{\hat{\mu}}\mu}$ is the background AdS vierbein. The anti-commutator is then
\beq
\{\gamma^{\mu},\gamma^{\nu}\}=2g^{\mu\nu}\,,
\eeq
with $g_{\mu\nu}$ the background AdS metric. We also define
\beq
\gamma^{\mu_1\cdots\mu_n}\equiv \gamma^{[\mu_1}\cdots\gamma^{\mu_n]}\,,
\eeq
and the matrix $\gamma_5\equiv i\gamma_{\hat{0}}\gamma_{\hat{1}}\gamma_{\hat{2}}\gamma_{\hat{3}}$, which satisfies
\beq
\{\gamma^{\mu},\gamma_5\}=0\,,\qquad (\gamma_5)^2=1\,.
\eeq

We use 4-component notation for spinors and omit the spinor indices throughout. The Majorana conjugate of a spinor $\chi$ is given by $\bar{\chi}=\chi^TC$, where the charge conjugation matrix $C$ satisfies $C^T=C^{\dagger}=C^{-1}=-C\,$. The charge conjugate is defined as $\chi^C=B^{-1}\chi^{*}$, where the matrix $B$ can be defined by the relation $(\gamma^{\mu})^{*}=B\gamma^{\mu}B^{-1}$\,. A Majorana spinor then satisfies the constraint $\chi^C=\chi\,$. In four dimensions there exists a so-called ``really real'' representation in which $B=1$, so that the gamma matrices are all real and the Majorana constraint simply states that the spinor is real.

When acting on spinors, the commutator of two AdS covariant derivatives yields an extra piece coming from the spin connection acting on the suppressed spinor index, for example
\beq\bal
\left[\nabla_{\mu},\nabla_{\nu}\right]\chi&=-\frac{1}{2L^2}\,\gamma_{\mu\nu}\chi\,,\\
\left[\nabla_{\mu},\nabla_{\nu}\right]\Psi_\rho&=-\frac{2}{L^2}\,g_{\rho[\mu}\Psi_{\nu]}-\frac{1}{2L^2}\,\gamma_{\mu\nu}\Psi_\rho\,.\\
\eal\eeq
The same is true for the Lie derivative, for instance
\beq\bal
{\mathcal L}_{\xi}\chi&=\xi^{\mu}\nabla_{\mu}\chi+\frac{1}{4}(\nabla_{\mu}\xi_{\nu})\gamma^{\mu\nu}\chi\,,\\
{\mathcal L}_{\xi}\Psi_{\mu}&=\xi^{\rho}\nabla_{\rho}\Psi_{\mu}+\nabla_{\mu}\xi^{\rho}\Psi_{\rho}+\frac{1}{4}(\nabla_{\rho}\xi_{\sigma})\gamma^{\rho\sigma}\Psi_{\mu}\,.\\
\eal\eeq

\section{Other explicit AdS super-multiplets} \label{app:other multiplets}

\subsection{AdS Wess--Zumino multiplet}

The free AdS Wess--Zumino multiplet $\left\{0\right\}_\Delta$, containing massive spins $(1/2,0,0)$ \cite{Breitenlohner:1982bm,deWit:1999ui} has the action
\beq
S=\int d^4x\sqrt{-g}\left[-\frac{1}{2}(\nabla\varphi)^2-\frac{1}{2}(\nabla\pi)^2-\frac{1}{2}\,\bar{\lambda}\gamma^{\mu}\nabla_{\mu}\lambda-\frac{m_0^2}{2}\,\varphi^2-\frac{m_0^{\prime2}}{2}\,\pi^2+\frac{m_{1/2}}{2}\,\bar{\lambda}\lambda\right]\,,
\eeq
with $\pi$ a pseudo-scalar. The scalar masses are given in terms of the spinor mass as
\beq \label{eq:wz multiplet masses}
m_0^2=m_{1/2}^2-\frac{m_{1/2}}{L}-\frac{2}{L^2}\,,\qquad m_0^{\prime2}=m_{1/2}^2+\frac{m_{1/2}}{L}-\frac{2}{L^2}\,,
\eeq
and the SUSY transformation rule reads
\beq\bal \label{eq:wz multiplet susy}
\delta_{\epsilon}\varphi&=\bar{\lambda}\epsilon\,,\qquad \delta_{\epsilon}\pi=\bar{\lambda}i\gamma_5\epsilon\,,\\
\delta_{\epsilon}\lambda&=\gamma^{\mu}(\nabla_{\mu}\varphi+i\gamma_5\nabla_{\mu}\pi)\epsilon+\left(m_{1/2}+\frac{1}{L}\right)\varphi\epsilon+\left(m_{1/2}-\frac{1}{L}\right)\pi i\gamma_5\epsilon\,.
\eal\eeq

\subsection{AdS gravitino multiplet\label{adsgravitinoappendix}}

The free AdS gravitino multiplet $\left\{1\right\}_\Delta$, containing massive spins $(3/2,1,1,1/2)$, was investigated in \cite{Zinoviev:2007ig} using a full St\"uckelberg formulation. Here we provide the result in unitary gauge. The action is
\beq\bal
S&=\int d^4x\sqrt{-g}\bigg[-\frac{1}{2}\,\bar{\Phi}_{\mu}\gamma^{\mu\nu\rho}\nabla_{\nu}\Phi_{\rho}-\frac{1}{4}\,F^{\mu\nu}F_{\mu\nu}-\frac{1}{4}\,G^{\mu\nu}G_{\mu\nu}-\frac{1}{2}\,\bar{\lambda}\gamma^{\mu}\nabla_{\mu}\lambda\\
&\quad+\frac{\mu}{2}\,\bar{\Phi}_{\mu}\gamma^{\mu\nu}\Phi_{\nu}-\frac{m_1^2}{2}\,A^{\mu}A_{\mu}-\frac{m_1^{\prime2}}{2}\,B^{\mu}B_{\mu}+\frac{m_{1/2}}{2}\,\bar{\lambda}\lambda\bigg]\,,
\eal\eeq
where $B_{\mu}$ is a pseudo-vector and the physical gravitino mass is given by $m_{3/2}^2=\mu^2-1/L^2\,$. The spinor and vector masses are determined in terms of the scale $\mu$ as
\beq \label{eq:gravitino multiplet masses}
m_{1/2}=\mu\,,\qquad m_1^2=\mu^2+\frac{\mu}{L}\,,\qquad m_1^{\prime2}=\mu^2-\frac{\mu}{L}\,,
\eeq
and the action is invariant under the supersymmetry
\beq\bal \label{eq:gravitino multiplet susy}
\delta_{\epsilon}\Phi_{\mu}&=\Big[-\frac{2\tilde{\mu}-3}{12}\,F_{\rho\sigma}\gamma^{\rho\sigma}\gamma_{\mu}-\frac{\tilde{\mu}}{3}\,\gamma^{\nu}(\nabla_{\mu}A_{\nu}+\nabla_{\nu}A_{\mu})\Big]\epsilon\\
&\quad+\frac{L}{6}\,\nabla_{\mu}\Big[F_{\nu\rho}\gamma^{\nu\rho}\epsilon\Big]+\frac{1}{L}\Big[\frac{2\tilde{\mu}(\tilde{\mu}-2)}{3}\,A_{\mu}-\frac{\tilde{\mu}(\tilde{\mu}-1)}{3}\,A^{\nu}\gamma_{\mu\nu}\Big]\epsilon\,,\\
&\quad+\sqrt{\frac{\tilde{\mu}-1}{\tilde{\mu}+1}}\bigg\{\Big[\frac{2\tilde{\mu}+3}{12}\,G_{\rho\sigma}\gamma^{\rho\sigma}\gamma_{\mu}i\gamma_5+\frac{\tilde{\mu}}{3}\,\gamma^{\nu}i\gamma_5(\nabla_{\mu}B_{\nu}+\nabla_{\nu}B_{\mu})\Big]\epsilon\\
&\quad-\frac{L}{6}\,\nabla_{\mu}\Big[G_{\nu\rho}\gamma^{\nu\rho}i\gamma_5\epsilon\Big]+\frac{1}{L}\Big[-\frac{2\tilde{\mu}(\tilde{\mu}+2)}{3}\,B_{\mu}i\gamma_5+\frac{\tilde{\mu}(\tilde{\mu}+1)}{3}\,B^{\nu}\gamma_{\mu\nu}i\gamma_5\Big]\epsilon\bigg\}\,,\\
\delta_{\epsilon} A_{\mu}&=(1-\tilde{\mu})\bar{\Phi}_{\mu}\epsilon+\frac{\sqrt{\tilde{\mu}^2-1}}{\sqrt{3}}\,\bar{\lambda}\gamma_{\mu}\epsilon+\frac{L}{\sqrt{3}}\sqrt{\frac{\tilde{\mu}-1}{\tilde{\mu}+1}}\,\nabla_{\mu}\left[\bar{\lambda}\epsilon\right]\,,\\
\delta_{\epsilon} B_{\mu}&=\sqrt{\tilde{\mu}^2-1}\,\bar{\Phi}_{\mu}i\gamma_5\epsilon-\frac{|\tilde{\mu}-1|}{\sqrt{3}}\,\bar{\lambda}\gamma_{\mu}i\gamma_5\epsilon-\frac{L}{\sqrt{3}}\,\frac{|\tilde{\mu}-1|}{\tilde{\mu}-1}\,\nabla_{\mu}\left[\bar{\lambda}i\gamma_5\epsilon\right]\,,\\
\delta_{\epsilon}\lambda&=\frac{\sqrt{\tilde{\mu}^2-1}}{2\sqrt{3}}\Big[F_{\mu\nu}\gamma^{\mu\nu}+\frac{2\tilde{\mu}}{L}\,A^{\mu}\gamma_{\mu}\Big]\epsilon-\frac{|\tilde{\mu}-1|}{2\sqrt{3}}\Big[G_{\mu\nu}\gamma^{\mu\nu}i\gamma_5+\frac{2\tilde{\mu}}{L}\,B^{\mu}\gamma_{\mu}i\gamma_5\Big]\epsilon\,,
\eal\eeq
with the shorthand notation $\tilde{\mu}\equiv \mu L\,$.

\bibliographystyle{utphys}
\addcontentsline{toc}{section}{References}
\bibliography{SUSYPMdraft_arxiv}

\providecommand{\href}[2]{#2}\begingroup\raggedright\begin{thebibliography}{10}

\bibitem{Deser:1983tm}
S.~Deser and R.~I. Nepomechie, ``{Anomalous Propagation of Gauge Fields in
  Conformally Flat Spaces},''
\href{http://dx.doi.org/10.1016/0370-2693(83)90317-9}{{\em Phys. Lett.}
  {\bfseries 132B} (1983) 321--324}.

\bibitem{Deser:1983mm}
S.~Deser and R.~I. Nepomechie, ``{Gauge Invariance Versus Masslessness in De
  Sitter Space},''
\href{http://dx.doi.org/10.1016/0003-4916(84)90156-8}{{\em Annals Phys.}
  {\bfseries 154} (1984) 396}.

\bibitem{Higuchi:1986py}
A.~Higuchi, ``{Forbidden Mass Range for Spin-2 Field Theory in De Sitter
  Space-time},''
\href{http://dx.doi.org/10.1016/0550-3213(87)90691-2}{{\em Nucl. Phys.}
  {\bfseries B282} (1987) 397--436}.

\bibitem{Brink:2000ag}
L.~Brink, R.~R. Metsaev, and M.~A. Vasiliev, ``{How massless are massless
  fields in AdS(d)},''
  \href{http://dx.doi.org/10.1016/S0550-3213(00)00402-8}{{\em Nucl. Phys.}
  {\bfseries B586} (2000) 183--205},
\href{http://arxiv.org/abs/hep-th/0005136}{{\ttfamily arXiv:hep-th/0005136
  [hep-th]}}.

\bibitem{Deser:2001pe}
S.~Deser and A.~Waldron, ``{Gauge invariances and phases of massive higher
  spins in (A)dS},''
  \href{http://dx.doi.org/10.1103/PhysRevLett.87.031601}{{\em Phys. Rev. Lett.}
  {\bfseries 87} (2001) 031601},
\href{http://arxiv.org/abs/hep-th/0102166}{{\ttfamily arXiv:hep-th/0102166
  [hep-th]}}.

\bibitem{Deser:2001us}
S.~Deser and A.~Waldron, ``{Partial masslessness of higher spins in (A)dS},''
  \href{http://dx.doi.org/10.1016/S0550-3213(01)00212-7}{{\em Nucl. Phys.}
  {\bfseries B607} (2001) 577--604},
\href{http://arxiv.org/abs/hep-th/0103198}{{\ttfamily arXiv:hep-th/0103198
  [hep-th]}}.

\bibitem{Deser:2001wx}
S.~Deser and A.~Waldron, ``{Stability of massive cosmological gravitons},''
  \href{http://dx.doi.org/10.1016/S0370-2693(01)00523-8}{{\em Phys. Lett.}
  {\bfseries B508} (2001) 347--353},
\href{http://arxiv.org/abs/hep-th/0103255}{{\ttfamily arXiv:hep-th/0103255
  [hep-th]}}.

\bibitem{Deser:2001xr}
S.~Deser and A.~Waldron, ``{Null propagation of partially massless higher spins
  in (A)dS and cosmological constant speculations},''
  \href{http://dx.doi.org/10.1016/S0370-2693(01)00756-0}{{\em Phys. Lett.}
  {\bfseries B513} (2001) 137--141},
\href{http://arxiv.org/abs/hep-th/0105181}{{\ttfamily arXiv:hep-th/0105181
  [hep-th]}}.

\bibitem{Zinoviev:2001dt}
{\relax Yu}.~M. Zinoviev, ``{On massive high spin particles in AdS},''
\href{http://arxiv.org/abs/hep-th/0108192}{{\ttfamily arXiv:hep-th/0108192
  [hep-th]}}.

\bibitem{Skvortsov:2006at}
E.~D. Skvortsov and M.~A. Vasiliev, ``{Geometric formulation for partially
  massless fields},''
  \href{http://dx.doi.org/10.1016/j.nuclphysb.2006.06.019}{{\em Nucl. Phys.}
  {\bfseries B756} (2006) 117--147},
\href{http://arxiv.org/abs/hep-th/0601095}{{\ttfamily arXiv:hep-th/0601095
  [hep-th]}}.

\bibitem{Skvortsov:2009zu}
E.~D. Skvortsov, ``{Gauge fields in (A)dS(d) and Connections of its symmetry
  algebra},'' \href{http://dx.doi.org/10.1088/1751-8113/42/38/385401}{{\em J.
  Phys.} {\bfseries A42} (2009) 385401},
\href{http://arxiv.org/abs/0904.2919}{{\ttfamily arXiv:0904.2919 [hep-th]}}.

\bibitem{Dolan:2008vc}
F.~A. Dolan, ``{On Superconformal Characters and Partition Functions in Three
  Dimensions},'' \href{http://dx.doi.org/10.1063/1.3211091}{{\em J. Math.
  Phys.} {\bfseries 51} (2010) 022301},
\href{http://arxiv.org/abs/0811.2740}{{\ttfamily arXiv:0811.2740 [hep-th]}}.

\bibitem{Bhattacharya:2008zy}
J.~Bhattacharya, S.~Bhattacharyya, S.~Minwalla, and S.~Raju, ``{Indices for
  Superconformal Field Theories in 3,5 and 6 Dimensions},''
  \href{http://dx.doi.org/10.1088/1126-6708/2008/02/064}{{\em JHEP} {\bfseries
  02} (2008) 064},
\href{http://arxiv.org/abs/0801.1435}{{\ttfamily arXiv:0801.1435 [hep-th]}}.

\bibitem{Cordova:2016emh}
C.~Cordova, T.~T. Dumitrescu, and K.~Intriligator, ``{Multiplets of
  Superconformal Symmetry in Diverse Dimensions},''
\href{http://arxiv.org/abs/1612.00809}{{\ttfamily arXiv:1612.00809 [hep-th]}}.

\bibitem{Oshima:2016gqy}
Y.~Oshima and M.~Yamazaki, ``{Determinant Formula for Parabolic Verma Modules
  of Lie Superalgebras},''
  \href{http://dx.doi.org/10.1016/j.jalgebra.2017.11.011}{{\em J. Algebra}
  {\bfseries 495} (2018) 51--80},
\href{http://arxiv.org/abs/1603.06705}{{\ttfamily arXiv:1603.06705 [math.RT]}}.

\bibitem{Sen:2018del}
K.~Sen and M.~Yamazaki, ``{Polology of Superconformal Blocks},''
\href{http://arxiv.org/abs/1810.01264}{{\ttfamily arXiv:1810.01264 [hep-th]}}.

\bibitem{Brust:2016gjy}
C.~Brust and K.~Hinterbichler, ``{Free $\square^{k}$ scalar conformal field
  theory},'' \href{http://dx.doi.org/10.1007/JHEP02(2017)066}{{\em JHEP}
  {\bfseries 02} (2017) 066},
\href{http://arxiv.org/abs/1607.07439}{{\ttfamily arXiv:1607.07439 [hep-th]}}.

\bibitem{Malaeb:2013lia}
O.~Malaeb, ``{Massive Gravity with $N=1$ local Supersymmetry},''
  \href{http://dx.doi.org/10.1140/epjc/s10052-013-2549-9}{{\em Eur. Phys. J.}
  {\bfseries C73} no.~9, (2013) 2549},
\href{http://arxiv.org/abs/1302.5092}{{\ttfamily arXiv:1302.5092 [hep-th]}}.

\bibitem{Malaeb:2013nra}
O.~Malaeb, ``{Supersymmetrizing Massive Gravity},''
  \href{http://dx.doi.org/10.1103/PhysRevD.88.025002}{{\em Phys. Rev.}
  {\bfseries D88} no.~2, (2013) 025002},
\href{http://arxiv.org/abs/1303.3580}{{\ttfamily arXiv:1303.3580 [hep-th]}}.

\bibitem{DelMonte:2016czb}
F.~Del~Monte, D.~Francia, and P.~A. Grassi, ``{Multimetric Supergravities},''
  \href{http://dx.doi.org/10.1007/JHEP09(2016)064}{{\em JHEP} {\bfseries 09}
  (2016) 064},
\href{http://arxiv.org/abs/1605.06793}{{\ttfamily arXiv:1605.06793 [hep-th]}}.

\bibitem{Ondo:2016cdv}
N.~A. Ondo and A.~J. Tolley, ``{Deconstructing Supergravity, I: Massive
  Supermultiplets},''
\href{http://arxiv.org/abs/1612.08752}{{\ttfamily arXiv:1612.08752 [hep-th]}}.

\bibitem{Zinoviev:2018juc}
Y.~M. Zinoviev, ``{On massive super(bi)gravity in the constructive approach},''
  \href{http://dx.doi.org/10.1088/1361-6382/aad1fb}{{\em Class. Quant. Grav.}
  {\bfseries 35} no.~17, (2018) 175006},
\href{http://arxiv.org/abs/1805.01650}{{\ttfamily arXiv:1805.01650 [hep-th]}}.

\bibitem{Zinoviev:2006im}
{\relax Yu}.~M. Zinoviev, ``{On massive spin 2 interactions},''
  \href{http://dx.doi.org/10.1016/j.nuclphysb.2007.02.005}{{\em Nucl. Phys.}
  {\bfseries B770} (2007) 83--106},
\href{http://arxiv.org/abs/hep-th/0609170}{{\ttfamily arXiv:hep-th/0609170
  [hep-th]}}.

\bibitem{Hassan:2012gz}
S.~F. Hassan, A.~Schmidt-May, and M.~von Strauss, ``{On Partially Massless
  Bimetric Gravity},''
  \href{http://dx.doi.org/10.1016/j.physletb.2013.09.021}{{\em Phys. Lett.}
  {\bfseries B726} (2013) 834--838},
\href{http://arxiv.org/abs/1208.1797}{{\ttfamily arXiv:1208.1797 [hep-th]}}.

\bibitem{Hassan:2012rq}
S.~F. Hassan, A.~Schmidt-May, and M.~von Strauss, ``{Bimetric theory and
  partial masslessness with Lanczos?Lovelock terms in arbitrary dimensions},''
  \href{http://dx.doi.org/10.1088/0264-9381/30/18/184010}{{\em Class. Quant.
  Grav.} {\bfseries 30} (2013) 184010},
\href{http://arxiv.org/abs/1212.4525}{{\ttfamily arXiv:1212.4525 [hep-th]}}.

\bibitem{deRham:2012kf}
C.~de~Rham and S.~Renaux-Petel, ``{Massive Gravity on de Sitter and Unique
  Candidate for Partially Massless Gravity},''
  \href{http://dx.doi.org/10.1088/1475-7516/2013/01/035}{{\em JCAP} {\bfseries
  1301} (2013) 035},
\href{http://arxiv.org/abs/1206.3482}{{\ttfamily arXiv:1206.3482 [hep-th]}}.

\bibitem{Hassan:2013pca}
S.~F. Hassan, A.~Schmidt-May, and M.~von Strauss, ``{Higher Derivative Gravity
  and Conformal Gravity From Bimetric and Partially Massless Bimetric
  Theory},'' \href{http://dx.doi.org/10.3390/universe1020092}{{\em Universe}
  {\bfseries 1} no.~2, (2015) 92--122},
\href{http://arxiv.org/abs/1303.6940}{{\ttfamily arXiv:1303.6940 [hep-th]}}.

\bibitem{Deser:2013uy}
S.~Deser, M.~Sandora, and A.~Waldron, ``{Nonlinear Partially Massless from
  Massive Gravity?},'' \href{http://dx.doi.org/10.1103/PhysRevD.87.101501}{{\em
  Phys. Rev.} {\bfseries D87} no.~10, (2013) 101501},
\href{http://arxiv.org/abs/1301.5621}{{\ttfamily arXiv:1301.5621 [hep-th]}}.

\bibitem{deRham:2013wv}
C.~de~Rham, K.~Hinterbichler, R.~A. Rosen, and A.~J. Tolley, ``{Evidence for
  and obstructions to nonlinear partially massless gravity},''
  \href{http://dx.doi.org/10.1103/PhysRevD.88.024003}{{\em Phys. Rev.}
  {\bfseries D88} no.~2, (2013) 024003},
\href{http://arxiv.org/abs/1302.0025}{{\ttfamily arXiv:1302.0025 [hep-th]}}.

\bibitem{Zinoviev:2014zka}
{\relax Yu}.~M. Zinoviev, ``{Massive spin-2 in the Fradkin?Vasiliev formalism.
  I. Partially massless case},''
  \href{http://dx.doi.org/10.1016/j.nuclphysb.2014.07.013}{{\em Nucl. Phys.}
  {\bfseries B886} (2014) 712--732},
\href{http://arxiv.org/abs/1405.4065}{{\ttfamily arXiv:1405.4065 [hep-th]}}.

\bibitem{Garcia-Saenz:2014cwa}
S.~Garcia-Saenz and R.~A. Rosen, ``{A non-linear extension of the spin-2
  partially massless symmetry},''
  \href{http://dx.doi.org/10.1007/JHEP05(2015)042}{{\em JHEP} {\bfseries 05}
  (2015) 042},
\href{http://arxiv.org/abs/1410.8734}{{\ttfamily arXiv:1410.8734 [hep-th]}}.

\bibitem{Hinterbichler:2014xga}
K.~Hinterbichler, ``{Manifest Duality Invariance for the Partially Massless
  Graviton},'' \href{http://dx.doi.org/10.1103/PhysRevD.91.026008}{{\em Phys.
  Rev.} {\bfseries D91} no.~2, (2015) 026008},
\href{http://arxiv.org/abs/1409.3565}{{\ttfamily arXiv:1409.3565 [hep-th]}}.

\bibitem{Joung:2014aba}
E.~Joung, W.~Li, and M.~Taronna, ``{No-Go Theorems for Unitary and Interacting
  Partially Massless Spin-Two Fields},''
  \href{http://dx.doi.org/10.1103/PhysRevLett.113.091101}{{\em Phys. Rev.
  Lett.} {\bfseries 113} (2014) 091101},
\href{http://arxiv.org/abs/1406.2335}{{\ttfamily arXiv:1406.2335 [hep-th]}}.

\bibitem{Alexandrov:2014oda}
S.~Alexandrov and C.~Deffayet, ``{On Partially Massless Theory in 3
  Dimensions},'' \href{http://dx.doi.org/10.1088/1475-7516/2015/03/043}{{\em
  JCAP} {\bfseries 1503} no.~03, (2015) 043},
\href{http://arxiv.org/abs/1410.2897}{{\ttfamily arXiv:1410.2897 [hep-th]}}.

\bibitem{Hassan:2015tba}
S.~F. Hassan, A.~Schmidt-May, and M.~von Strauss, ``{Extended Weyl Invariance
  in a Bimetric Model and Partial Masslessness},''
  \href{http://dx.doi.org/10.1088/0264-9381/33/1/015011}{{\em Class. Quant.
  Grav.} {\bfseries 33} no.~1, (2016) 015011},
\href{http://arxiv.org/abs/1507.06540}{{\ttfamily arXiv:1507.06540 [hep-th]}}.

\bibitem{Hinterbichler:2015nua}
K.~Hinterbichler and R.~A. Rosen, ``{Partially Massless Monopoles and
  Charges},'' \href{http://dx.doi.org/10.1103/PhysRevD.92.105019}{{\em Phys.
  Rev.} {\bfseries D92} no.~10, (2015) 105019},
\href{http://arxiv.org/abs/1507.00355}{{\ttfamily arXiv:1507.00355 [hep-th]}}.

\bibitem{Cherney:2015jxp}
D.~Cherney, S.~Deser, A.~Waldron, and G.~Zahariade, ``{Non-linear duality
  invariant partially massless models?},''
  \href{http://dx.doi.org/10.1016/j.physletb.2015.12.029}{{\em Phys. Lett.}
  {\bfseries B753} (2016) 293--296},
\href{http://arxiv.org/abs/1511.01053}{{\ttfamily arXiv:1511.01053 [hep-th]}}.

\bibitem{Gwak:2015vfb}
S.~Gwak, E.~Joung, K.~Mkrtchyan, and S.-J. Rey, ``{Rainbow Valley of Colored
  (Anti) de Sitter Gravity in Three Dimensions},''
  \href{http://dx.doi.org/10.1007/JHEP04(2016)055}{{\em JHEP} {\bfseries 04}
  (2016) 055},
\href{http://arxiv.org/abs/1511.05220}{{\ttfamily arXiv:1511.05220 [hep-th]}}.

\bibitem{Gwak:2015jdo}
S.~Gwak, E.~Joung, K.~Mkrtchyan, and S.-J. Rey, ``{Rainbow vacua of colored
  higher-spin (A)dS$_{3}$ gravity},''
  \href{http://dx.doi.org/10.1007/JHEP05(2016)150}{{\em JHEP} {\bfseries 05}
  (2016) 150},
\href{http://arxiv.org/abs/1511.05975}{{\ttfamily arXiv:1511.05975 [hep-th]}}.

\bibitem{Garcia-Saenz:2015mqi}
S.~Garcia-Saenz, K.~Hinterbichler, A.~Joyce, E.~Mitsou, and R.~A. Rosen,
  ``{No-go for Partially Massless Spin-2 Yang-Mills},''
  \href{http://dx.doi.org/10.1007/JHEP02(2016)043}{{\em JHEP} {\bfseries 02}
  (2016) 043},
\href{http://arxiv.org/abs/1511.03270}{{\ttfamily arXiv:1511.03270 [hep-th]}}.

\bibitem{Hinterbichler:2016fgl}
K.~Hinterbichler and A.~Joyce, ``{Manifest Duality for Partially Massless
  Higher Spins},'' \href{http://dx.doi.org/10.1007/JHEP09(2016)141}{{\em JHEP}
  {\bfseries 09} (2016) 141},
\href{http://arxiv.org/abs/1608.04385}{{\ttfamily arXiv:1608.04385 [hep-th]}}.

\bibitem{Bonifacio:2016blz}
J.~Bonifacio and K.~Hinterbichler, ``{Kaluza-Klein reduction of massive and
  partially massless spin-2 fields},''
  \href{http://dx.doi.org/10.1103/PhysRevD.95.024023}{{\em Phys. Rev.}
  {\bfseries D95} no.~2, (2017) 024023},
\href{http://arxiv.org/abs/1611.00362}{{\ttfamily arXiv:1611.00362 [hep-th]}}.

\bibitem{Apolo:2016ort}
L.~Apolo and S.~F. Hassan, ``{Non-linear partially massless symmetry in an
  SO(1,5) continuation of conformal gravity},''
  \href{http://dx.doi.org/10.1088/1361-6382/aa69f7}{{\em Class. Quant. Grav.}
  {\bfseries 34} no.~10, (2017) 105005},
\href{http://arxiv.org/abs/1609.09514}{{\ttfamily arXiv:1609.09514 [hep-th]}}.

\bibitem{Apolo:2016vkn}
L.~Apolo, S.~F. Hassan, and A.~Lundkvist, ``{Gauge and global symmetries of the
  candidate partially massless bimetric gravity},''
  \href{http://dx.doi.org/10.1103/PhysRevD.94.124055}{{\em Phys. Rev.}
  {\bfseries D94} no.~12, (2016) 124055},
\href{http://arxiv.org/abs/1609.09515}{{\ttfamily arXiv:1609.09515 [hep-th]}}.

\bibitem{Bernard:2017tcg}
L.~Bernard, C.~Deffayet, K.~Hinterbichler, and M.~von Strauss, ``{Partially
  Massless Graviton on Beyond Einstein Spacetimes},''
  \href{http://dx.doi.org/10.1103/PhysRevD.95.124036}{{\em Phys. Rev.}
  {\bfseries D95} no.~12, (2017) 124036},
\href{http://arxiv.org/abs/1703.02538}{{\ttfamily arXiv:1703.02538 [hep-th]}}.

\bibitem{Boulanger:2018dau}
N.~Boulanger, C.~Deffayet, S.~Garcia-Saenz, and L.~Traina, ``{Consistent
  deformations of free massive field theories in the Stueckelberg
  formulation},'' \href{http://dx.doi.org/10.1007/JHEP07(2018)021}{{\em JHEP}
  {\bfseries 07} (2018) 021},
\href{http://arxiv.org/abs/1806.04695}{{\ttfamily arXiv:1806.04695 [hep-th]}}.

\bibitem{Bekaert:2013zya}
X.~Bekaert and M.~Grigoriev, ``{Higher order singletons, partially massless
  fields and their boundary values in the ambient approach},''
  \href{http://dx.doi.org/10.1016/j.nuclphysb.2013.08.015}{{\em Nucl. Phys.}
  {\bfseries B876} (2013) 667--714},
\href{http://arxiv.org/abs/1305.0162}{{\ttfamily arXiv:1305.0162 [hep-th]}}.

\bibitem{Basile:2014wua}
T.~Basile, X.~Bekaert, and N.~Boulanger, ``{Flato-Fronsdal theorem for
  higher-order singletons},''
  \href{http://dx.doi.org/10.1007/JHEP11(2014)131}{{\em JHEP} {\bfseries 11}
  (2014) 131},
\href{http://arxiv.org/abs/1410.7668}{{\ttfamily arXiv:1410.7668 [hep-th]}}.

\bibitem{Alkalaev:2014nsa}
K.~B. Alkalaev, M.~Grigoriev, and E.~D. Skvortsov, ``{Uniformizing higher-spin
  equations},'' \href{http://dx.doi.org/10.1088/1751-8113/48/1/015401}{{\em J.
  Phys.} {\bfseries A48} no.~1, (2015) 015401},
\href{http://arxiv.org/abs/1409.6507}{{\ttfamily arXiv:1409.6507 [hep-th]}}.

\bibitem{Joung:2015jza}
E.~Joung and K.~Mkrtchyan, ``{Partially-massless higher-spin algebras and their
  finite-dimensional truncations},''
  \href{http://dx.doi.org/10.1007/JHEP01(2016)003}{{\em JHEP} {\bfseries 01}
  (2016) 003},
\href{http://arxiv.org/abs/1508.07332}{{\ttfamily arXiv:1508.07332 [hep-th]}}.

\bibitem{Brust:2016zns}
C.~Brust and K.~Hinterbichler, ``{Partially Massless Higher-Spin Theory},''
  \href{http://dx.doi.org/10.1007/JHEP02(2017)086}{{\em JHEP} {\bfseries 02}
  (2017) 086},
\href{http://arxiv.org/abs/1610.08510}{{\ttfamily arXiv:1610.08510 [hep-th]}}.

\bibitem{Maassarani:1996jn}
Z.~Maassarani and D.~Serban, ``{Nonunitary conformal field theory and
  logarithmic operators for disordered systems},''
  \href{http://dx.doi.org/10.1016/S0550-3213(97)00014-X}{{\em Nucl. Phys.}
  {\bfseries B489} (1997) 603--625},
\href{http://arxiv.org/abs/hep-th/9605062}{{\ttfamily arXiv:hep-th/9605062
  [hep-th]}}.

\bibitem{Penedones:2015aga}
J.~Penedones, E.~Trevisani, and M.~Yamazaki, ``{Recursion Relations for
  Conformal Blocks},'' \href{http://dx.doi.org/10.1007/JHEP09(2016)070}{{\em
  JHEP} {\bfseries 09} (2016) 070},
\href{http://arxiv.org/abs/1509.00428}{{\ttfamily arXiv:1509.00428 [hep-th]}}.

\bibitem{Carroll:2004st}
S.~M. Carroll, {\em {Spacetime and geometry: An introduction to general
  relativity}}.
\newblock 2004.
\newblock
\url{http://www.slac.stanford.edu/spires/find/books/www?cl=QC6:C37:2004}.
\newblock

\bibitem{Klebanov:1999tb}
I.~R. Klebanov and E.~Witten, ``{AdS / CFT correspondence and symmetry
  breaking},'' \href{http://dx.doi.org/10.1016/S0550-3213(99)00387-9}{{\em
  Nucl. Phys.} {\bfseries B556} (1999) 89--114},
\href{http://arxiv.org/abs/hep-th/9905104}{{\ttfamily arXiv:hep-th/9905104
  [hep-th]}}.

\bibitem{Mack:1975je}
G.~Mack, ``{All unitary ray representations of the conformal group SU(2,2) with
  positive energy},''
\href{http://dx.doi.org/10.1007/BF01613145}{{\em Commun. Math. Phys.}
  {\bfseries 55} (1977) 1}.

\bibitem{Jantzen1977}
J.~C. Jantzen, ``Kontravariante formen auf induzierten darstellungen
  halbeinfacher lie-algebren,''
  \href{http://dx.doi.org/10.1007/BF01391218}{{\em Mathematische Annalen}
  {\bfseries 226} no.~1, (Feb, 1977) 53--65}.
  \url{https://doi.org/10.1007/BF01391218}.

\bibitem{Minwalla:1997ka}
S.~Minwalla, ``{Restrictions imposed by superconformal invariance on quantum
  field theories},'' \href{http://dx.doi.org/10.4310/ATMP.1998.v2.n4.a4}{{\em
  Adv. Theor. Math. Phys.} {\bfseries 2} (1998) 783--851},
\href{http://arxiv.org/abs/hep-th/9712074}{{\ttfamily arXiv:hep-th/9712074
  [hep-th]}}.

\bibitem{Dolan:2001ih}
L.~Dolan, C.~R. Nappi, and E.~Witten, ``{Conformal operators for partially
  massless states},''
  \href{http://dx.doi.org/10.1088/1126-6708/2001/10/016}{{\em JHEP} {\bfseries
  10} (2001) 016},
\href{http://arxiv.org/abs/hep-th/0109096}{{\ttfamily arXiv:hep-th/0109096
  [hep-th]}}.

\bibitem{deWit:1999ui}
B.~de~Wit and I.~Herger, ``{Anti-de Sitter supersymmetry},'' {\em Lect. Notes
  Phys.} {\bfseries 541} (2000) 79--100,
  \href{http://arxiv.org/abs/hep-th/9908005}{{\ttfamily arXiv:hep-th/9908005
  [hep-th]}}.
[,79(1999)].

\bibitem{Simmons-Duffin:2016gjk}
D.~Simmons-Duffin, \href{http://dx.doi.org/10.1142/9789813149441_0001}{``{The
  Conformal Bootstrap},''} in {\em {Proceedings, Theoretical Advanced Study
  Institute in Elementary Particle Physics: New Frontiers in Fields and Strings
  (TASI 2015): Boulder, CO, USA, June 1-26, 2015}}, pp.~1--74.
\newblock 2017.
\newblock \href{http://arxiv.org/abs/1602.07982}{{\ttfamily arXiv:1602.07982
  [hep-th]}}.
\newblock
\url{https://inspirehep.net/record/1424282/files/arXiv:1602.07982.pdf}.
\newblock

\bibitem{Basile:2018eac}
T.~Basile, X.~Bekaert, and E.~Joung, ``{Conformal Higher-Spin Gravity:
  Linearized Spectrum = Symmetry Algebra},''
\href{http://arxiv.org/abs/1808.07728}{{\ttfamily arXiv:1808.07728 [hep-th]}}.

\bibitem{Dirac:1963ta}
P.~A.~M. Dirac, ``{A Remarkable representation of the 3 + 2 de Sitter group},''
\href{http://dx.doi.org/10.1063/1.1704016}{{\em J. Math. Phys.} {\bfseries 4}
  (1963) 901--909}.

\bibitem{Balasubramanian:1998sn}
V.~Balasubramanian, P.~Kraus, and A.~E. Lawrence, ``{Bulk versus boundary
  dynamics in anti-de Sitter space-time},''
  \href{http://dx.doi.org/10.1103/PhysRevD.59.046003}{{\em Phys. Rev.}
  {\bfseries D59} (1999) 046003},
\href{http://arxiv.org/abs/hep-th/9805171}{{\ttfamily arXiv:hep-th/9805171
  [hep-th]}}.

\bibitem{deRham:2018svs}
C.~De~Rham, K.~Hinterbichler, and L.~Johnson, ``{On the (A)dS Decoupling Limits
  of Massive Gravity},''
\href{http://arxiv.org/abs/1807.08754}{{\ttfamily arXiv:1807.08754 [hep-th]}}.

\bibitem{Buchbinder:2002gh}
I.~L. Buchbinder, S.~J. Gates, Jr., W.~D. Linch, III, and J.~Phillips, ``{New
  4-D, N=1 superfield theory: Model of free massive superspin 3/2 multiplet},''
  \href{http://dx.doi.org/10.1016/S0370-2693(02)01772-0}{{\em Phys. Lett.}
  {\bfseries B535} (2002) 280--288},
\href{http://arxiv.org/abs/hep-th/0201096}{{\ttfamily arXiv:hep-th/0201096
  [hep-th]}}.

\bibitem{Zinoviev:2002xn}
{\relax Yu}.~M. Zinoviev, ``{Massive spin two supermultiplets},''
\href{http://arxiv.org/abs/hep-th/0206209}{{\ttfamily arXiv:hep-th/0206209
  [hep-th]}}.

\bibitem{Haag:1974qh}
R.~Haag, J.~T. Lopuszanski, and M.~Sohnius, ``{All Possible Generators of
  Supersymmetries of the s Matrix},''
  \href{http://dx.doi.org/10.1016/0550-3213(75)90279-5}{{\em Nucl. Phys.}
  {\bfseries B88} (1975) 257}.
[,257(1974)].

\bibitem{Freedman:2012zz}
D.~Z. Freedman and A.~Van~Proeyen, {\em {Supergravity}}.
\newblock Cambridge Univ. Press, Cambridge, UK, 2012.
\newblock
\url{http://www.cambridge.org/mw/academic/subjects/physics/theoretical-physics-and-mathematical-physics/supergravity?format=AR}.
\newblock

\bibitem{Breitenlohner:1982bm}
P.~Breitenlohner and D.~Z. Freedman, ``{Positive Energy in anti-De Sitter
  Backgrounds and Gauged Extended Supergravity},''
\href{http://dx.doi.org/10.1016/0370-2693(82)90643-8}{{\em Phys. Lett.}
  {\bfseries 115B} (1982) 197--201}.

\bibitem{Zinoviev:2007ig}
{\relax Yu}.~M. Zinoviev, ``{Massive supermultiplets with spin 3/2},''
  \href{http://dx.doi.org/10.1088/1126-6708/2007/05/092}{{\em JHEP} {\bfseries
  05} (2007) 092},
\href{http://arxiv.org/abs/hep-th/0703118}{{\ttfamily arXiv:hep-th/0703118
  [hep-th]}}.

\end{thebibliography}\endgroup

\end{document}